\documentclass[useAMS,usenatbib]{mn2e}
\usepackage{graphicx}
\usepackage{booktabs}
\usepackage{amsmath}
\usepackage{amssymb}
\usepackage{threeparttable}
\usepackage[T1]{fontenc}
\usepackage{aecompl}

\title[Super flares on KIC10524994 \& KIC07133671?]{Superflares on the slowly rotating solar type stars KIC10524994 \& KIC07133671?}
\author[M. Kitze et al.]{M. Kitze$^{1}$\thanks{E-mail:
manfred.kitze@uni-jena.de}, R. Neuh\"auser$^1$, V. Hambaryan$^1$, C. Ginski$^1$\\
$^{1}$Astrophysikalisches Institut und Universit\"atssternwarte, Jena 07743, Germany}
\begin{document}

\date{Accepted ...}

\pagerange{\pageref{firstpage}--\pageref{lastpage}} \pubyear{2014}

\maketitle

\label{firstpage}

\begin{abstract}
{\bf An investigation of the G-type stellar population with Kepler \citep{maehara} shows, that less than $1\%$ of those stars show superflares. Due to the large pixel scale of Kepler ($\approx 4''/px$) it is still not clear whether the detected superflares really occur on the G-type stars. Knowing the origin of such large brightenings is important to study their frequency statistics, which are uncertain due to the low number of sun-like stars ($T_{eff}=5600-6000\:$K and $P_{rot} > 10\:$d) which are currently considered to exhibit superflares. We present a complete Kepler data analysis of the sun-like stars KIC10524994 and KIC07133671 (the only two stars within this subsample of solar twins with flare energies larger than $10^{35}\:$erg \citep{maehara}), regarding superflare properties and a study about their origin. We could detect 4 new superflares within the epoch \citet{maehara} investigated and found 14 superflares in the remaining light curve for KIC10524994. Astrometric Kepler data of KIC07133671 show that the photocenter is shifted by $0.006\:$px or $25\:$mas during the one detected flare. Hence the flare probably originated from another star directed towards the north east. This lowers the superflare rate of sun-like stars (and hence the Sun) for $E > 10^{35}\:$erg, since this additional star is probably not solar-like.}
\end{abstract}


\section{Introduction}

Stellar flares are defined as rapid brightenings of emission across the electromagnetic spectrum \citep{Benz}, typically followed by an exponential relaxation back to the quiescent state of a star. It is assumed that such brightenings are caused by the release of magnetic energy, which is stored near active regions in the stellar photosphere. Our Sun shows flares with typical energies of $10^{24}-10^{29}\:$erg \citep{schrijver}. A study of the spectral content of solar flares \citep{kretzschmar} indicates that the optical part of the spectrum can be well approximated by a black body radiation with $\sim 10000\:$K.

A variety of stars show flares with energies $\geq\:10$ times that of the largest known solar event ($10^{32}\:$erg), that are therefore called "superflares" \citep{schaefer}. It is expected that superflares occur more frequently on magnetically active K- and M-type dwarfs, which are mostly fast rotators and partly full convective \citep{maehara}. Nevertheless there is a small number of stars, presumably like the Sun, that show superflares. These stars are on or near the main sequence, have spectral types F8-G8 with corresponding temperatures $5100\:$K$\:\leq T_{eff}<6000\:$K, and are not young. A study of all the suitable stars within the Kepler Input Catalogue (KIC) \citep{brown} shows that the fraction of superflare stars among such "solar like" stars is about $0.5\:\%$ \citep{maehara}. This value must be treated with care, since flares can also be caused by close binary interaction, wide late-type companions or other scenarios. If such superflare stars are really solar twins, one might conclude upon the frequency and energetics of superflares on the Sun.

Solar superflares with energies above $10^{35}\:$erg could be harmful to the biosphere and also very relevant for space weather. It was discussed recently whether the AD 774/5 $^{14}$C event was due to a solar superflare \citep{miyake}. Since the activity of the Sun as known from historic sunspot and aurora observations is known only for up to 2-3 millennia, the rate of solar superflares is also not well-known. In Intcal $^{14}C$ data, there  were three large spikes within the last $3000\:$yr \citep{miyake},but their origin is not known. Better constraints can be obtained from the rate of superflares of many sun-like stars. 

In this work we reanalyzed the light curves of two interesting superflare stars, KIC10524994 and KIC07133671, found by \cite{maehara}, that have the highest flare energies within the subsample of presumably sun-like stars with temperatures $5600\:$K$\:\leq T_{eff}< 6000\:$K, and probably long rotational periods ($> 10\:$d). We investigated the origin of the superflares directly from Kepler Photometry (Sec. 2.1) and Kepler Astrometry (Sec. 2.5) and indirectly from the occurrence times of the detected flares (Sec. 2.3). Additionally we characterized all detected flares regarding their energies and luminosities (Sec. 2.4), determined a power law for the frequency of superflares $> 10^{34}\:$erg (Sec. 2.6) and we estimated the ages of the stars for a comparison with the Sun (Sec. 2.2).
\begin{figure*}
\includegraphics[angle=270,scale=0.19]{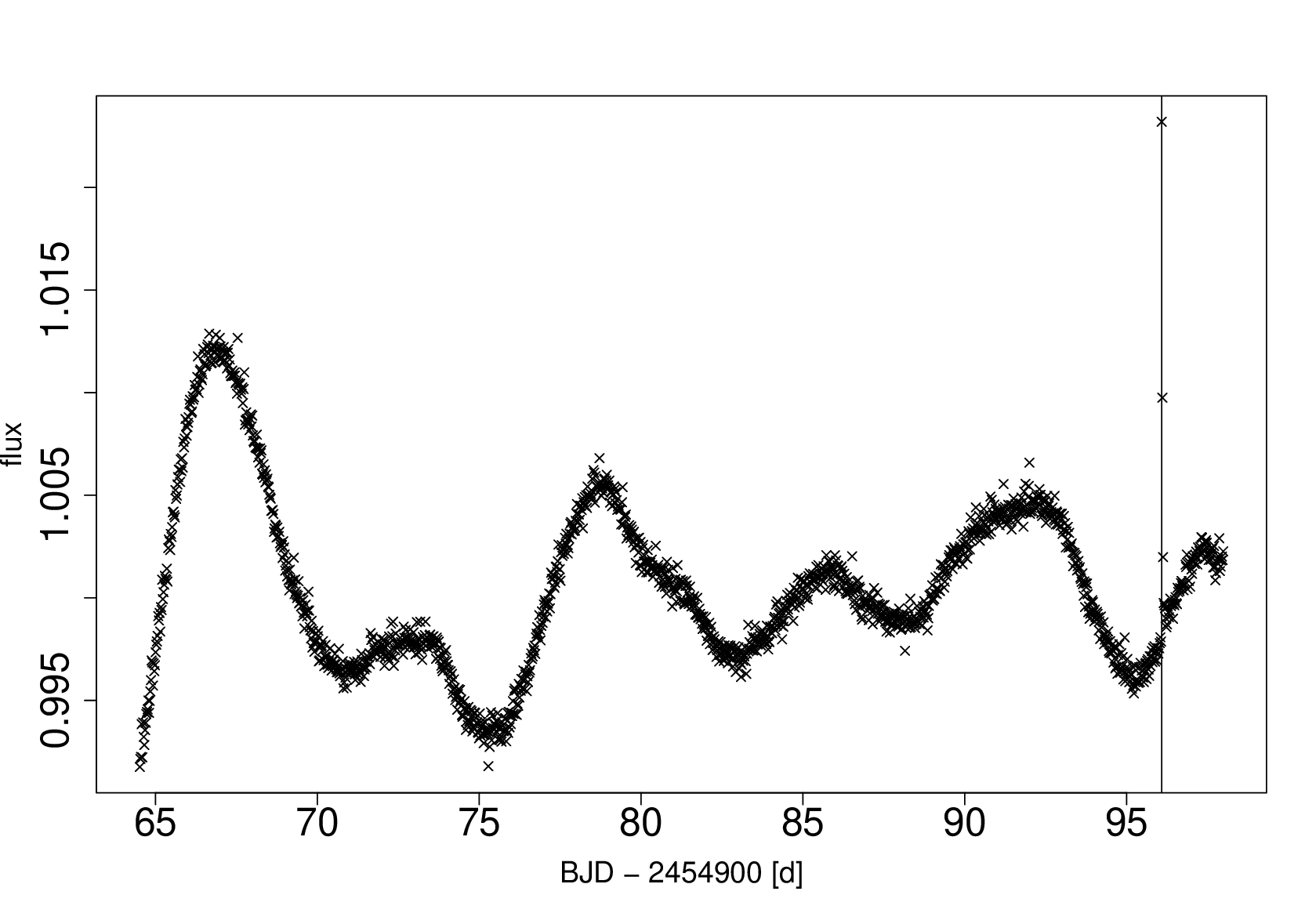}
\includegraphics[angle=270,scale=0.19]{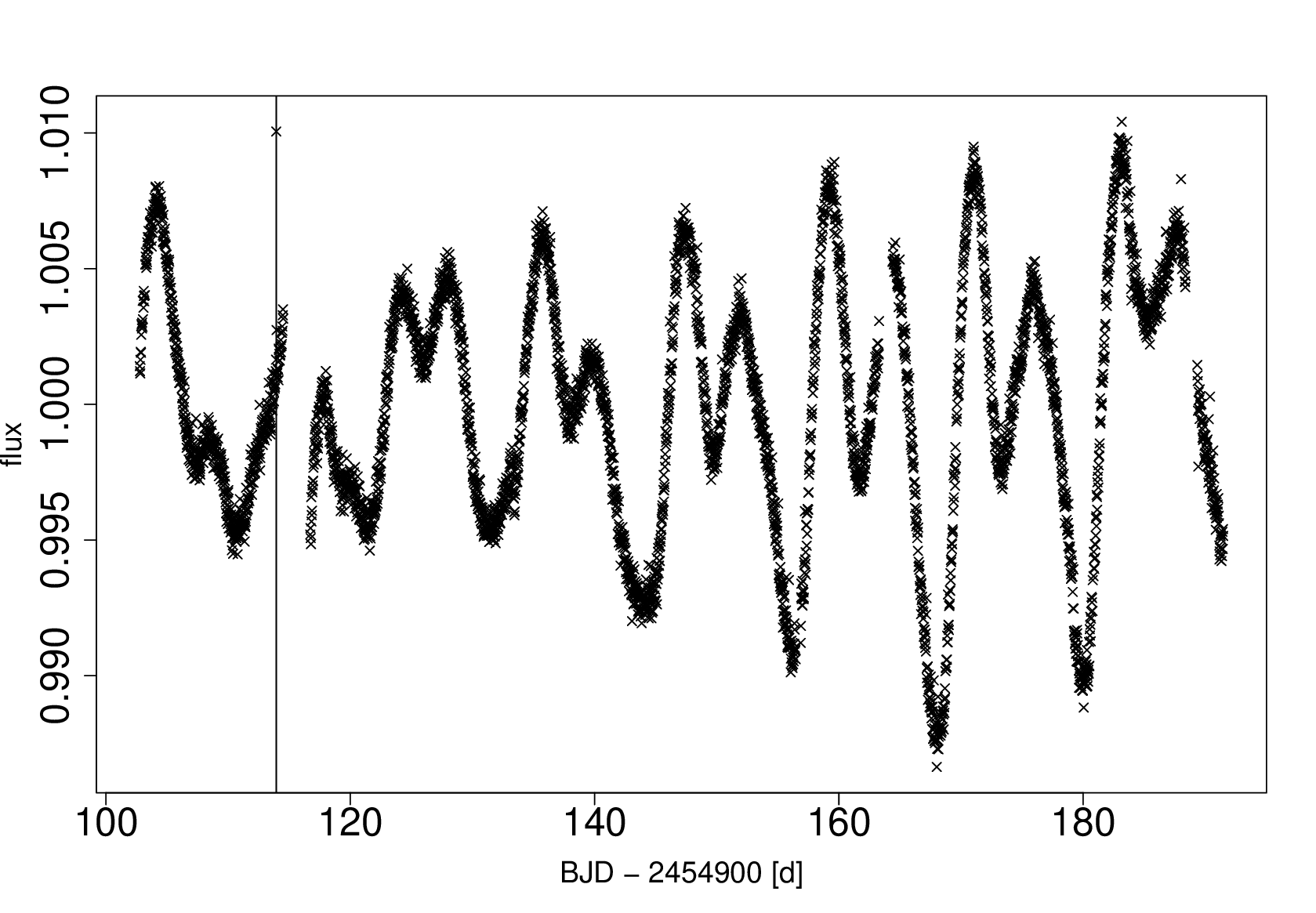}
\includegraphics[angle=270,scale=0.19]{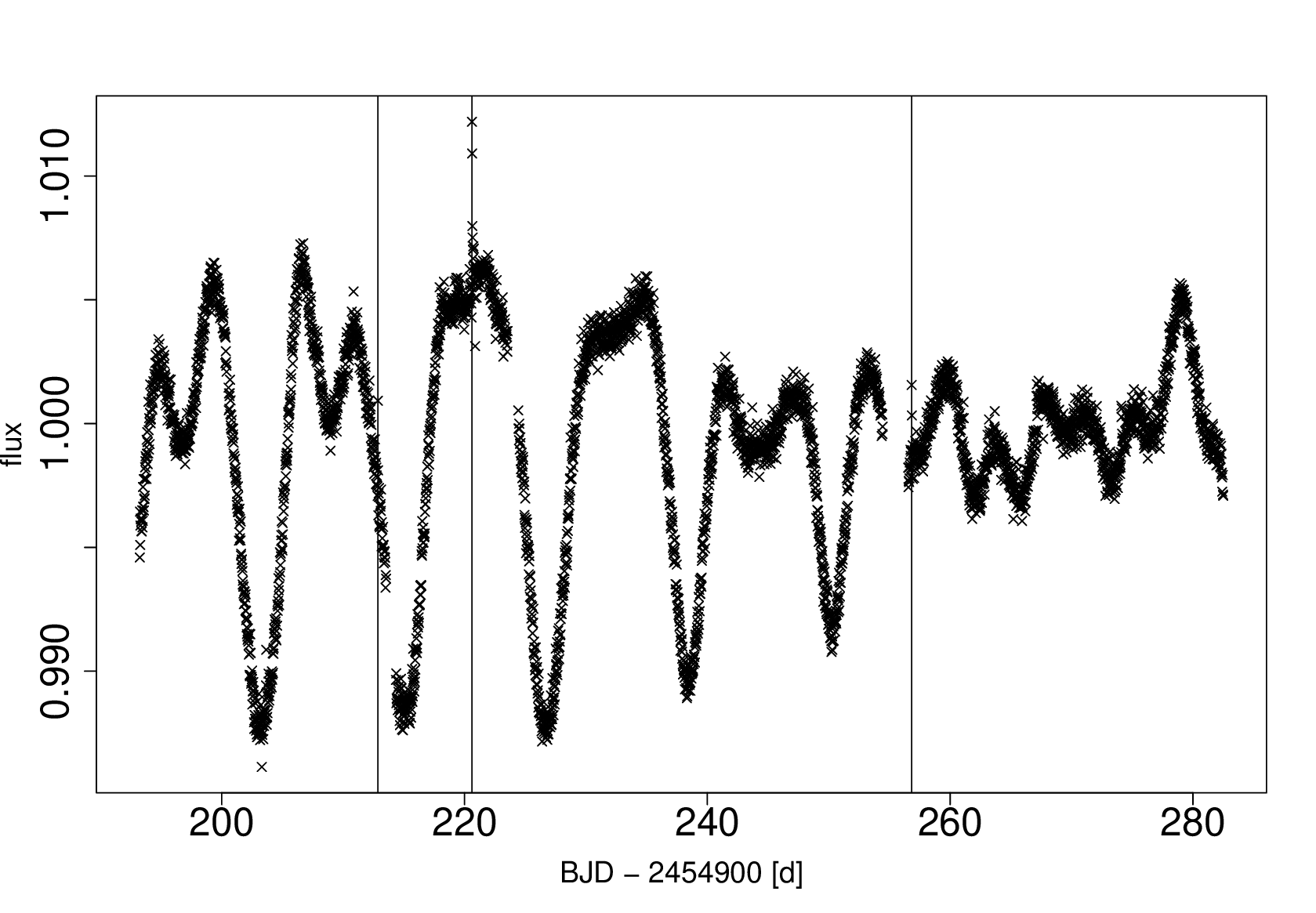}
\includegraphics[angle=270,scale=0.19]{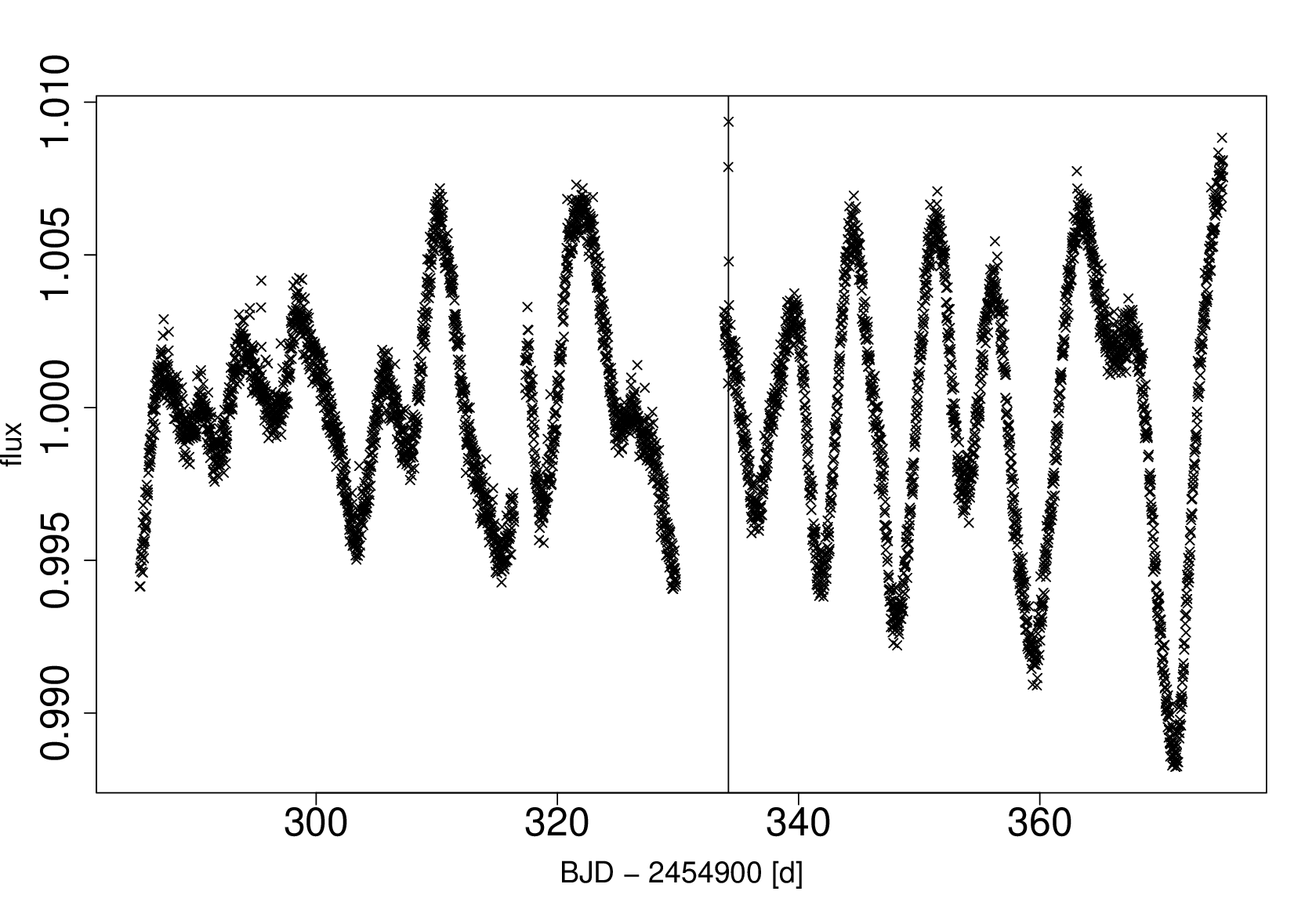}
\includegraphics[angle=270,scale=0.19]{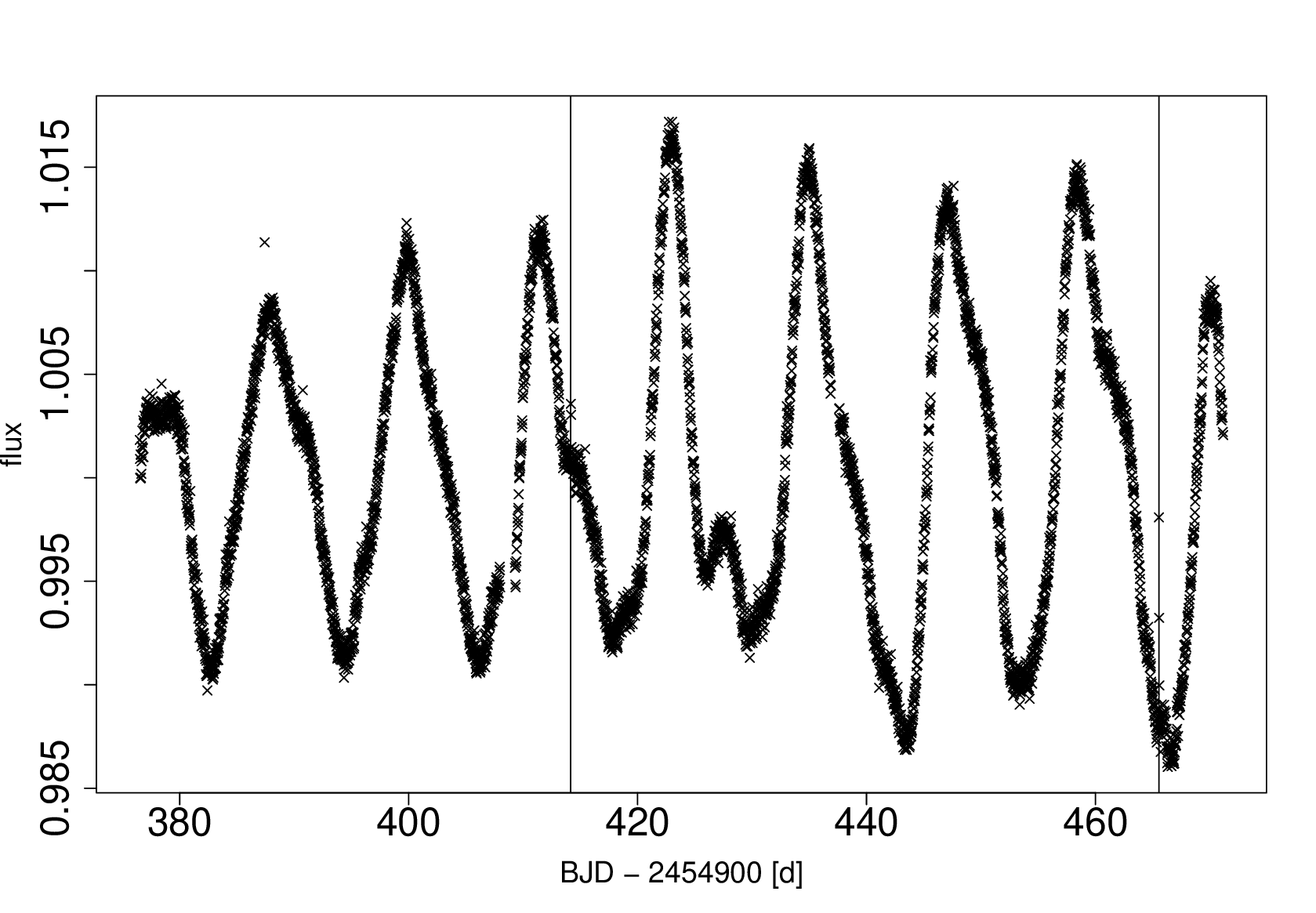}
\includegraphics[angle=270,scale=0.19]{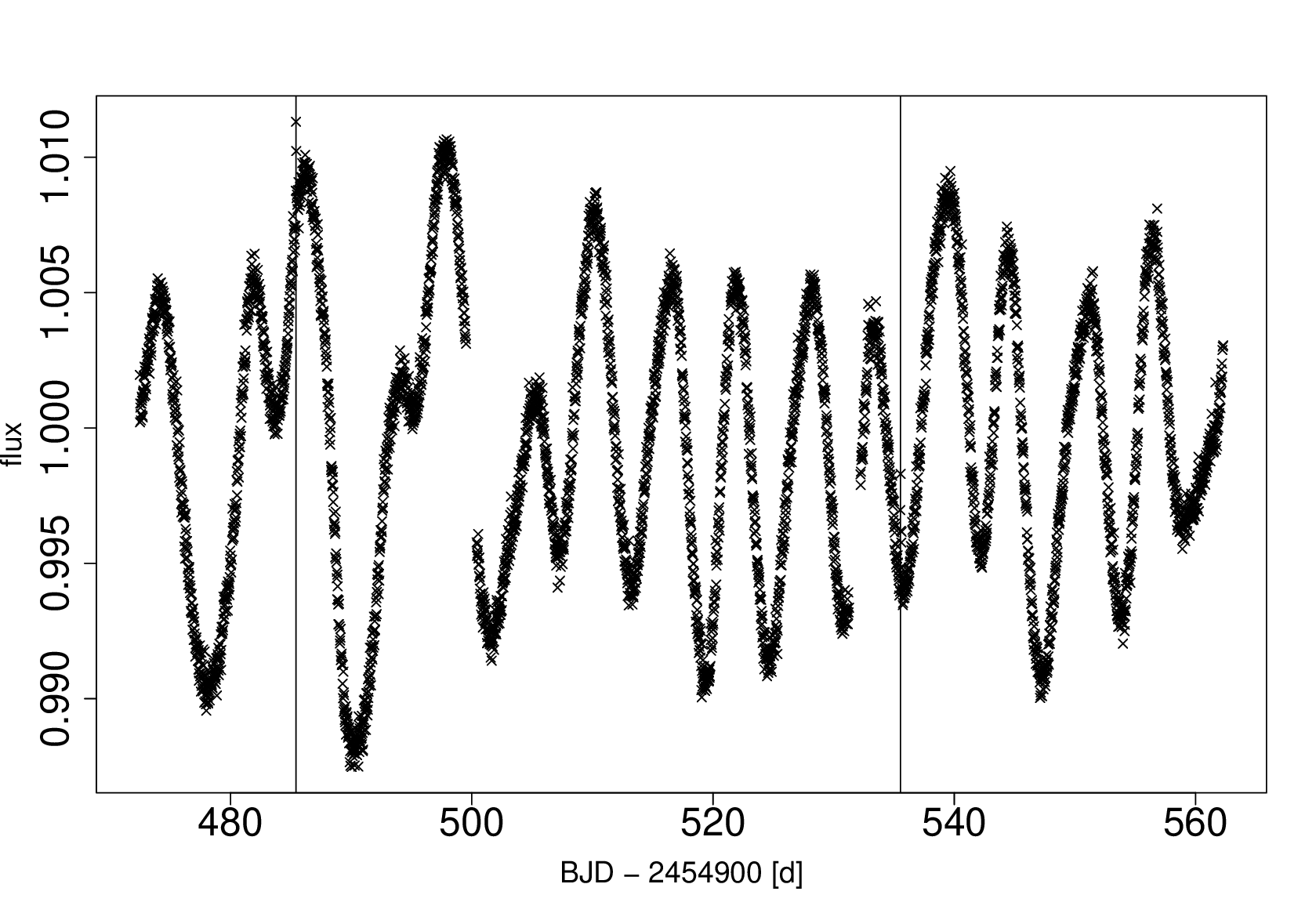}
\includegraphics[angle=270,scale=0.19]{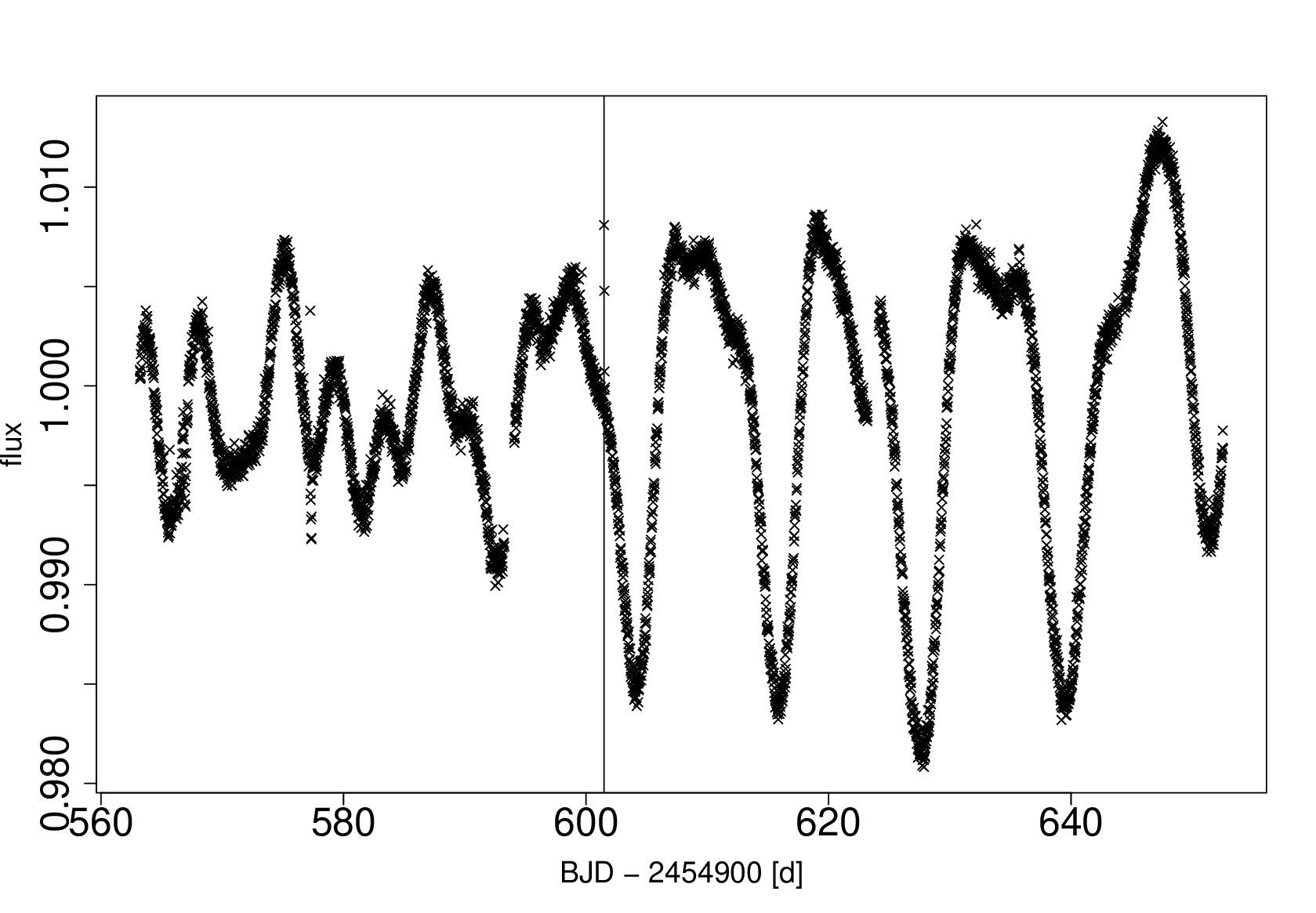}
\includegraphics[angle=270,scale=0.19]{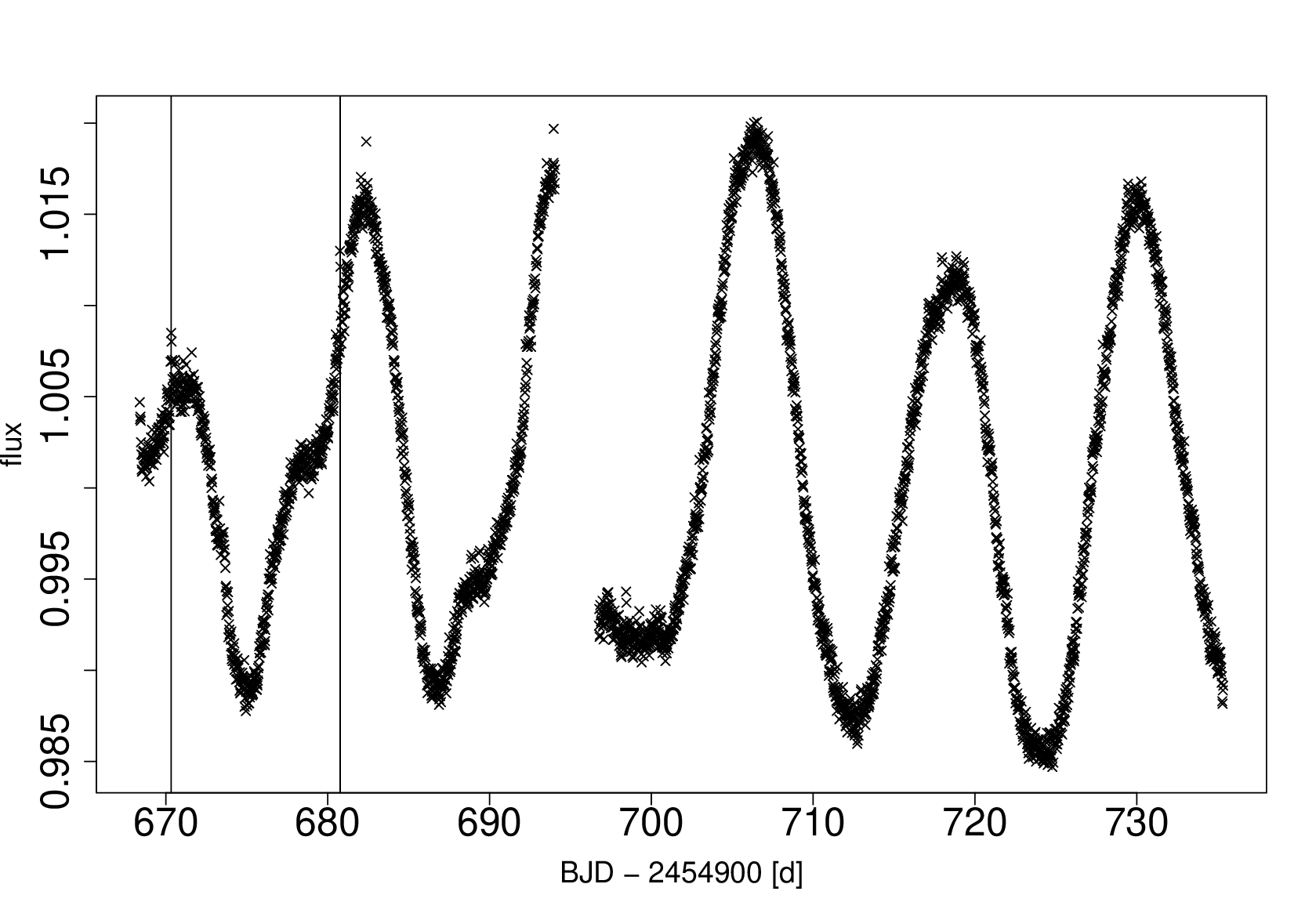}
\includegraphics[angle=270,scale=0.19]{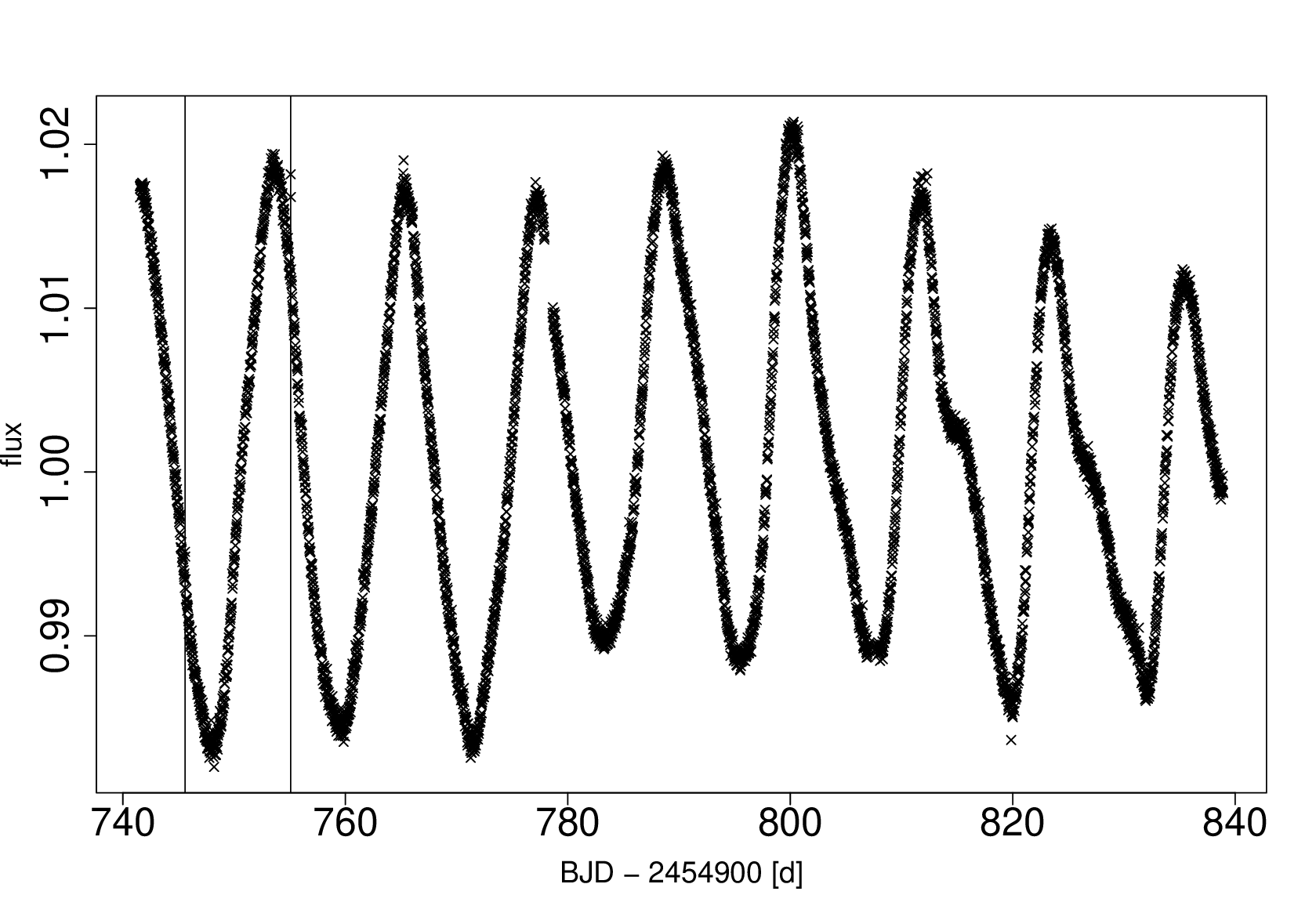}
\includegraphics[angle=270,scale=0.19]{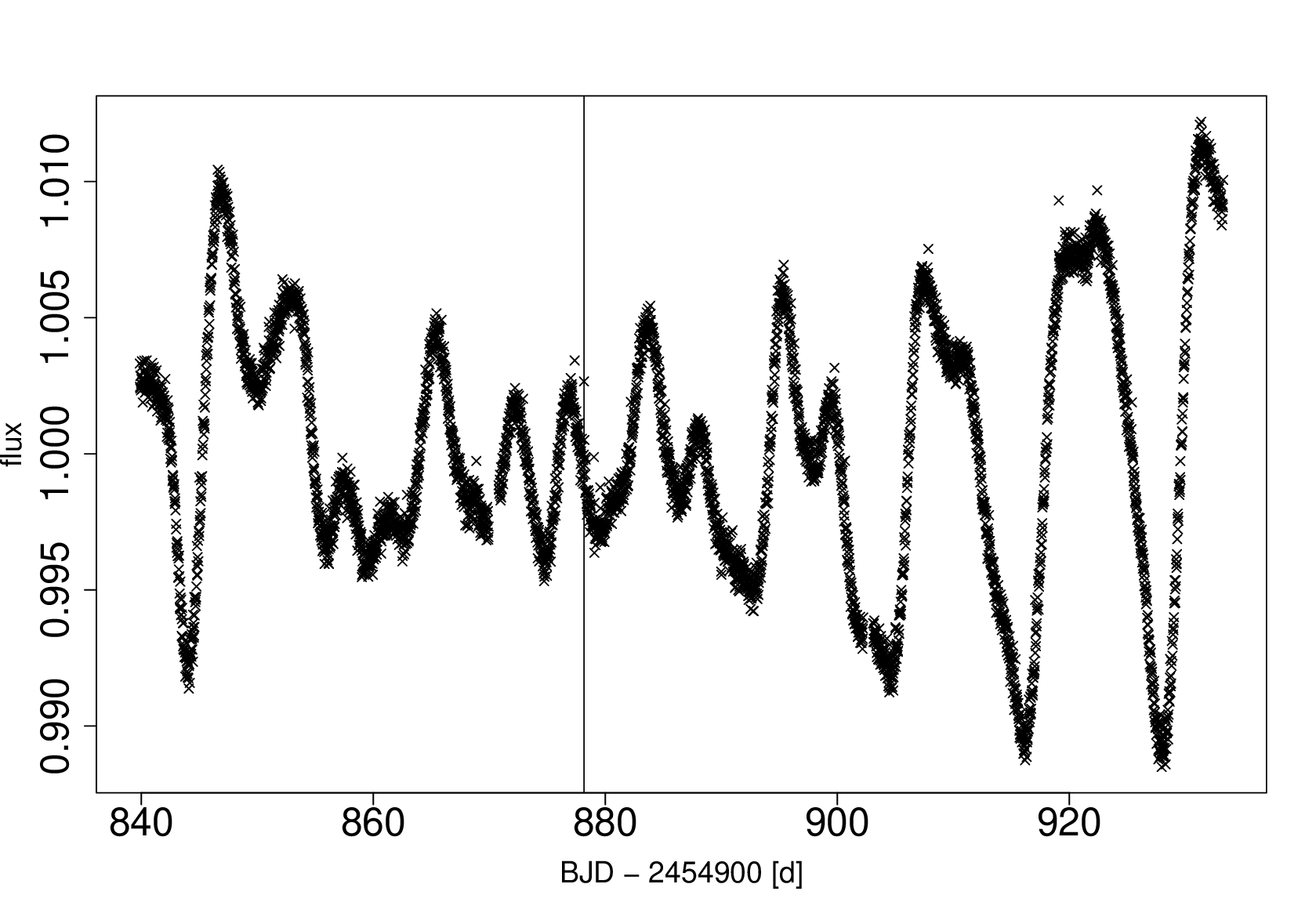}
\includegraphics[angle=270,scale=0.19]{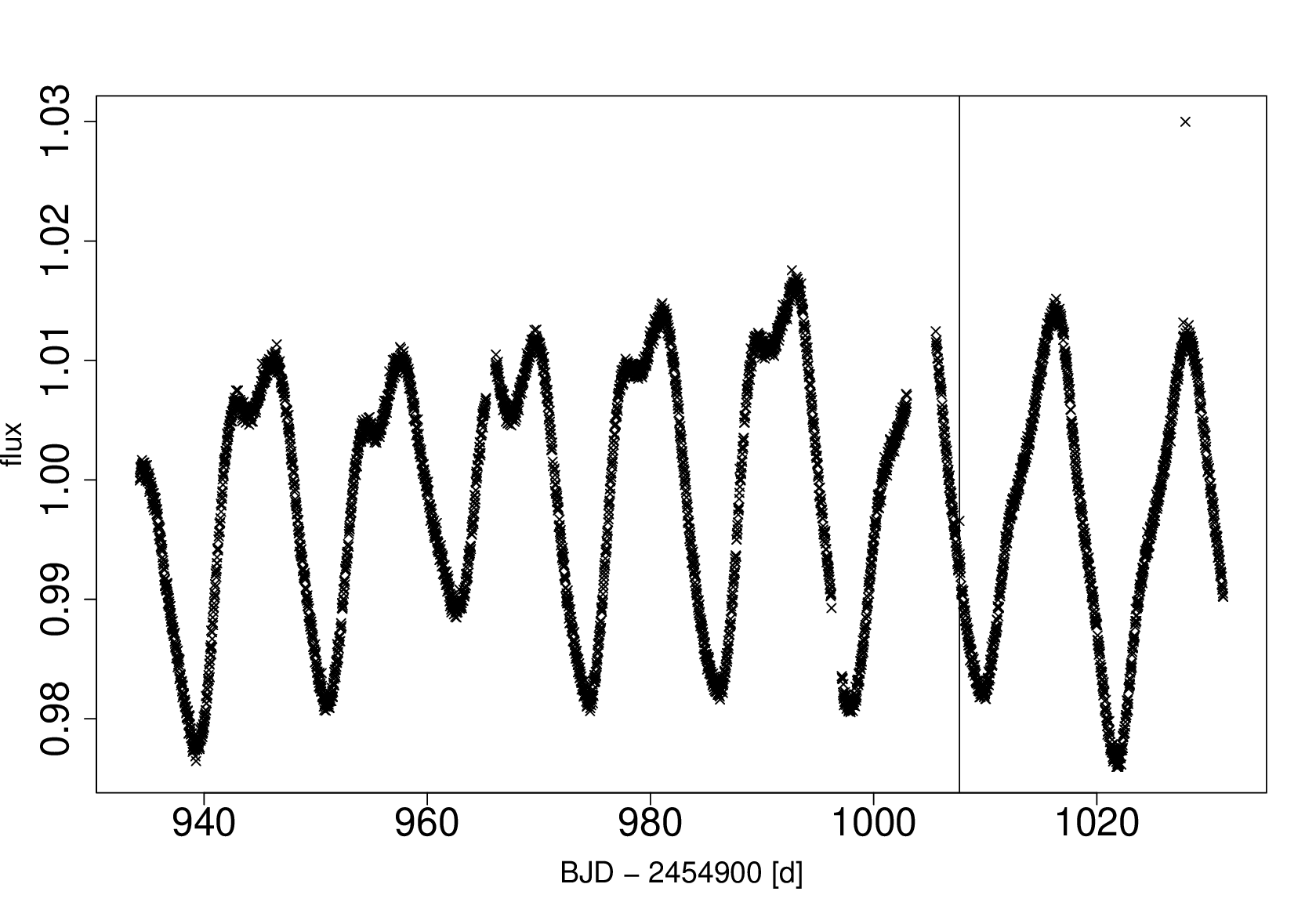}
\includegraphics[angle=270,scale=0.19]{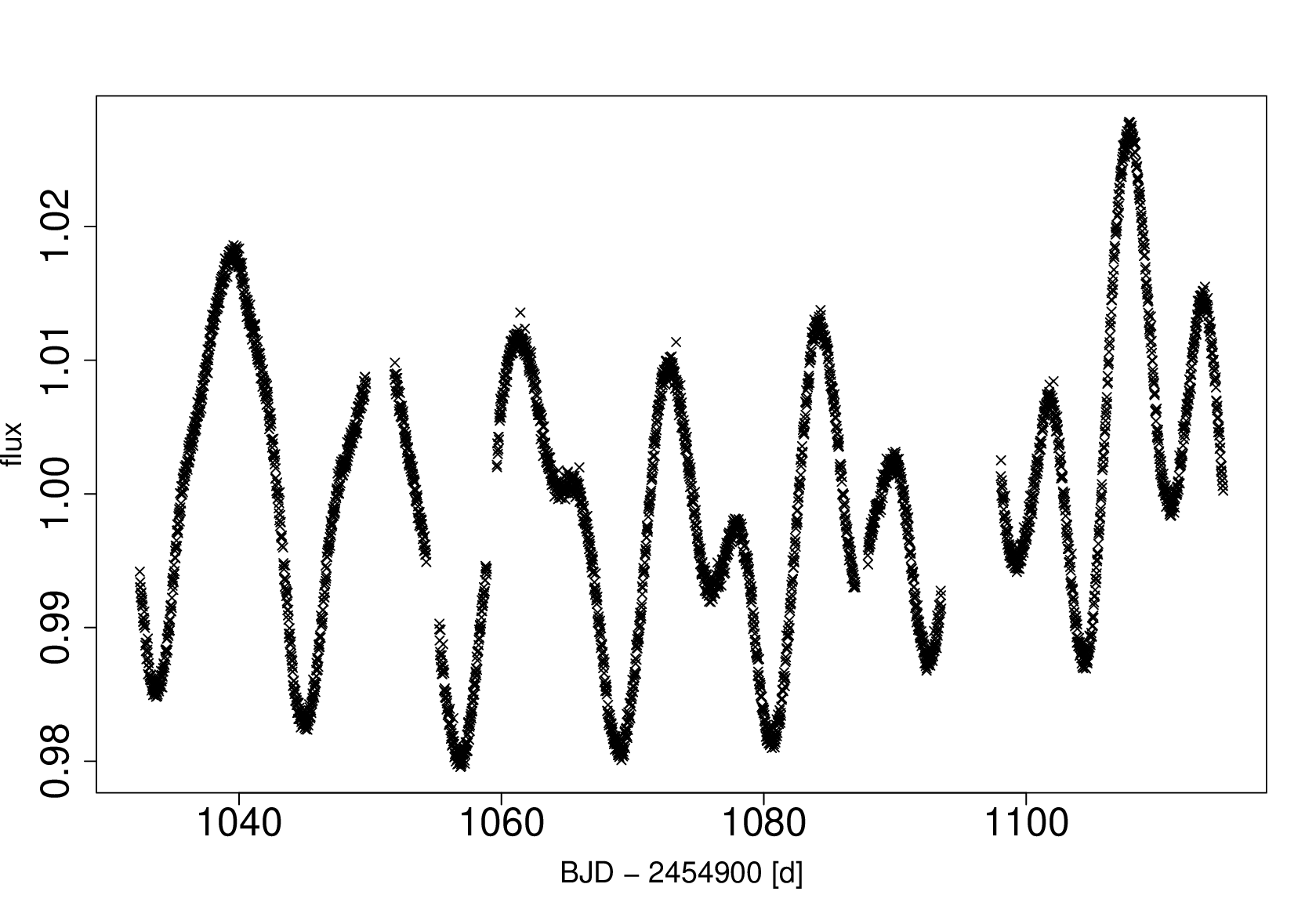}
\includegraphics[angle=270,scale=0.19]{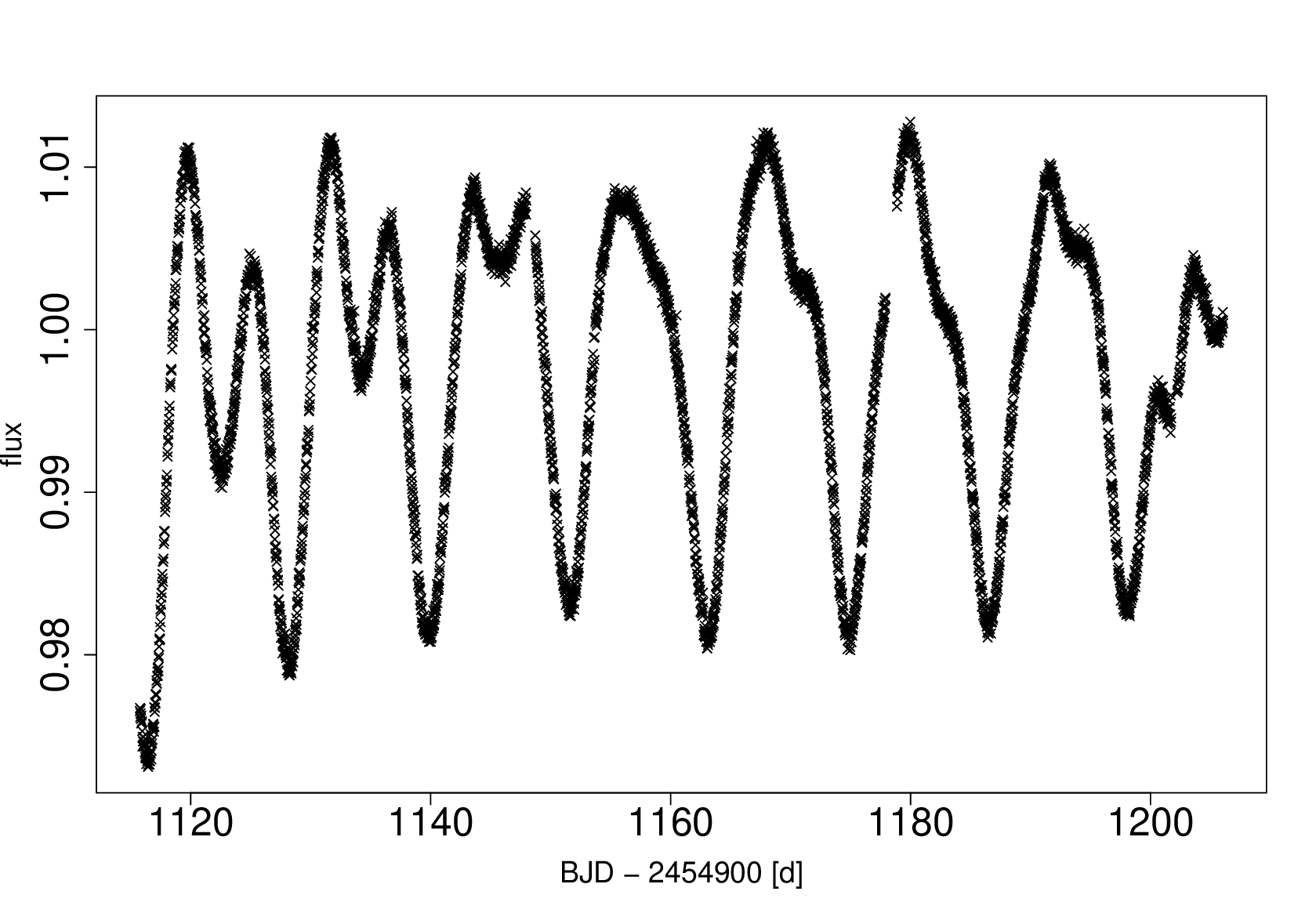}
\includegraphics[angle=270,scale=0.19]{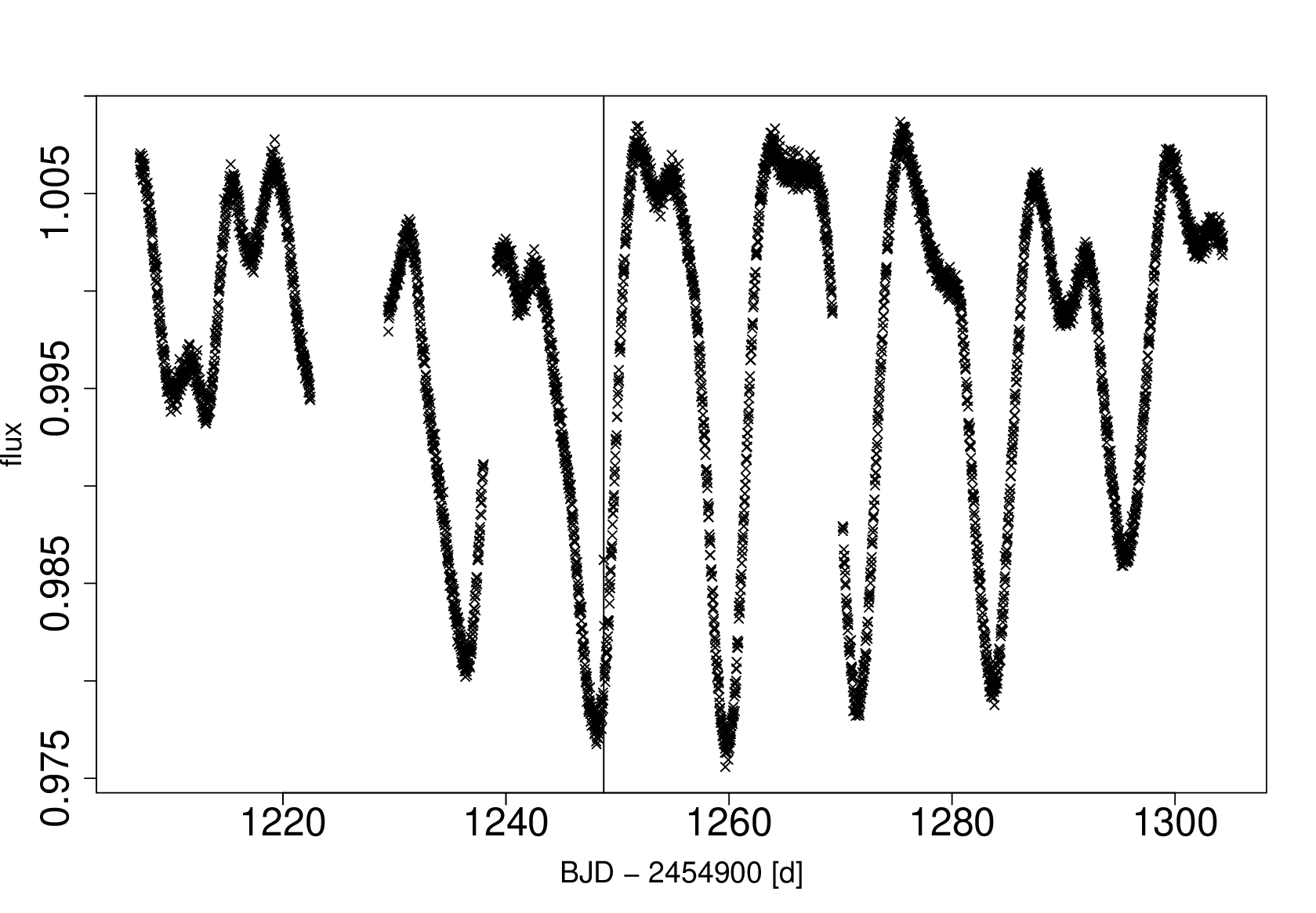}
\includegraphics[angle=270,scale=0.19]{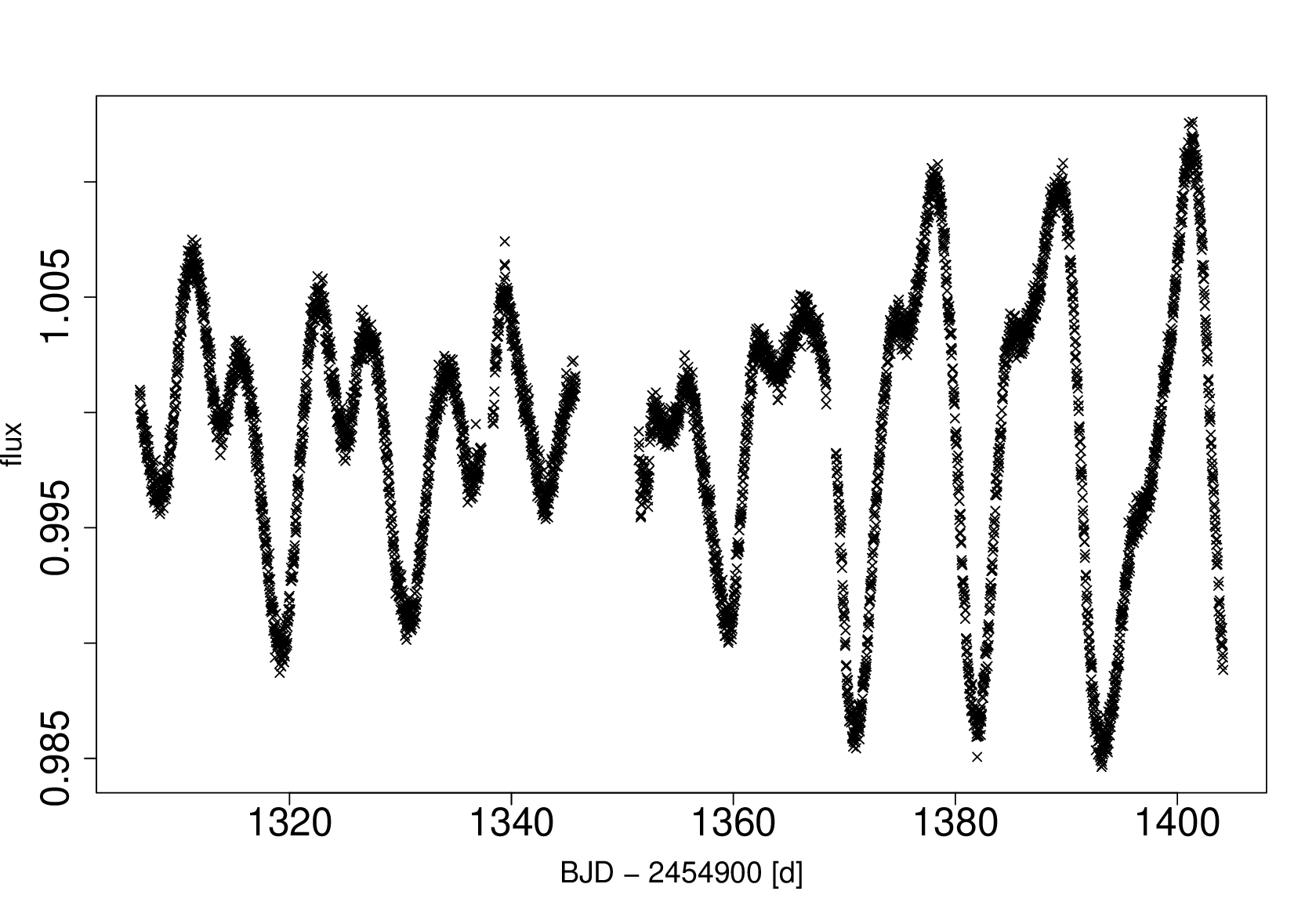}
\includegraphics[angle=270,scale=0.19]{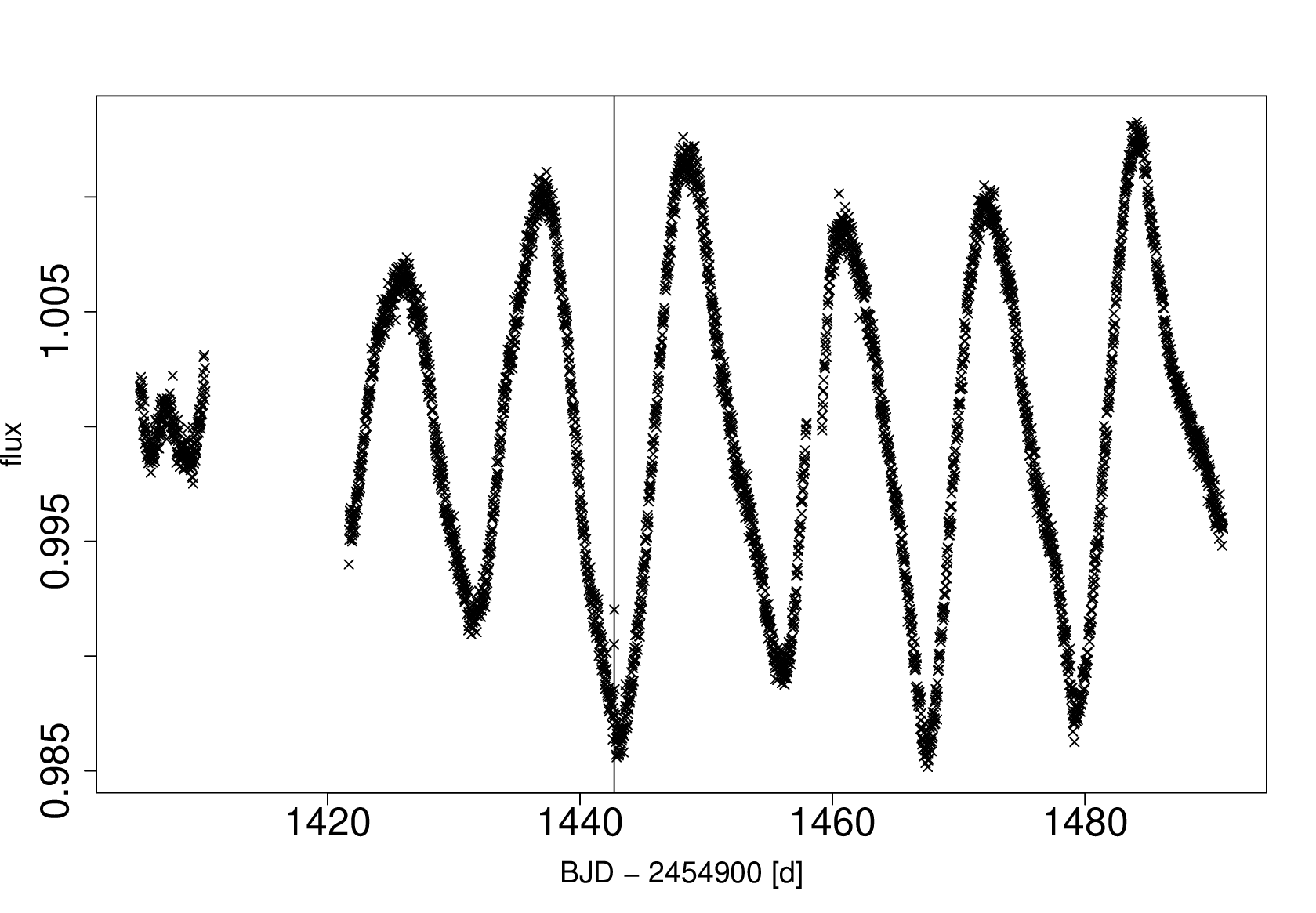}
 %\vspace{pt}
 \caption{Quarter light curves of KIC10524994 for the first 16 quarters of public Kepler data. Vertical lines denote the positions of detected flares (see Tab. \ref{properties}).}
  \label{curves_1}
\end{figure*}
\begin{figure*}
\includegraphics[angle=270,scale=0.19]{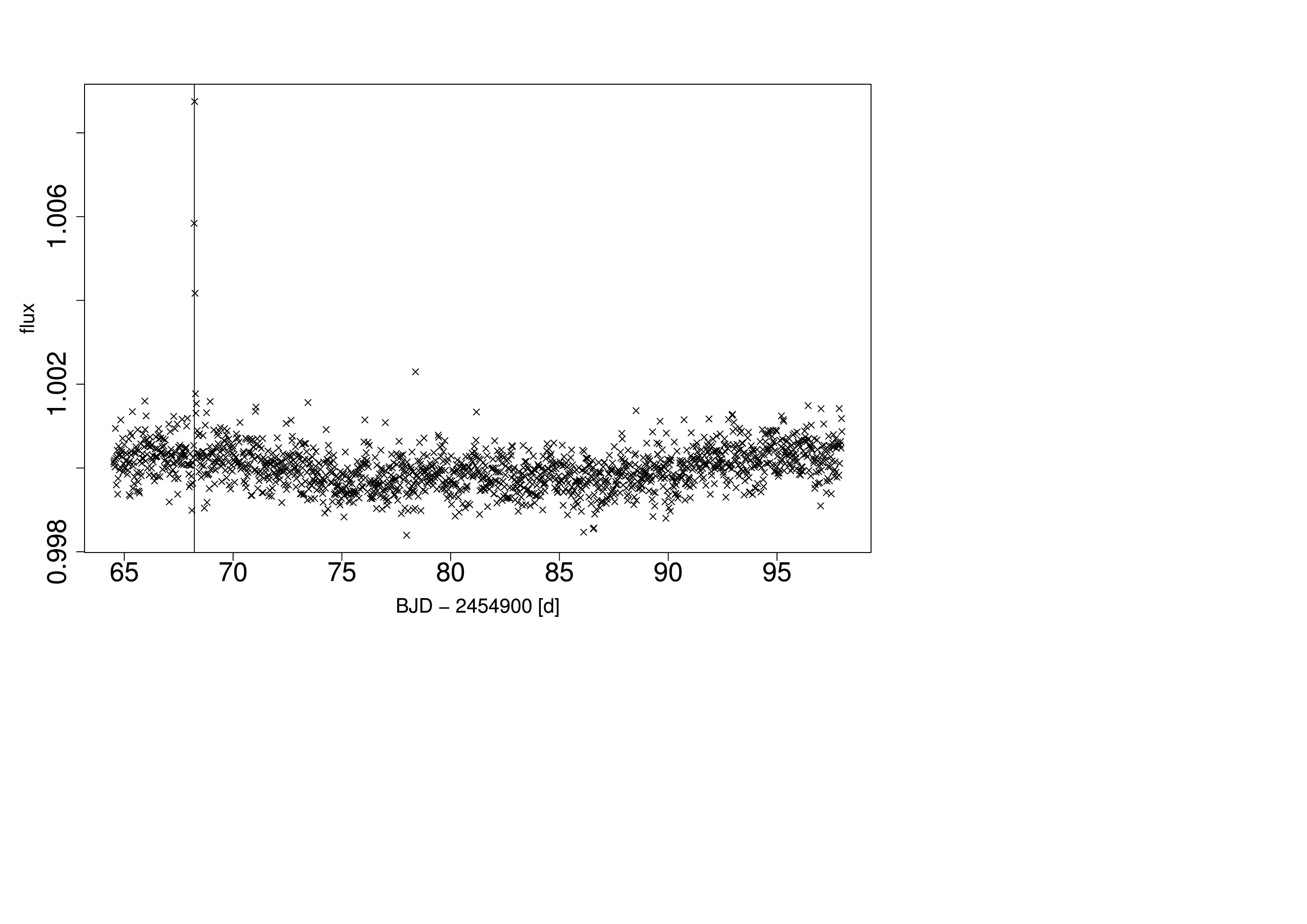}
\includegraphics[angle=270,scale=0.19]{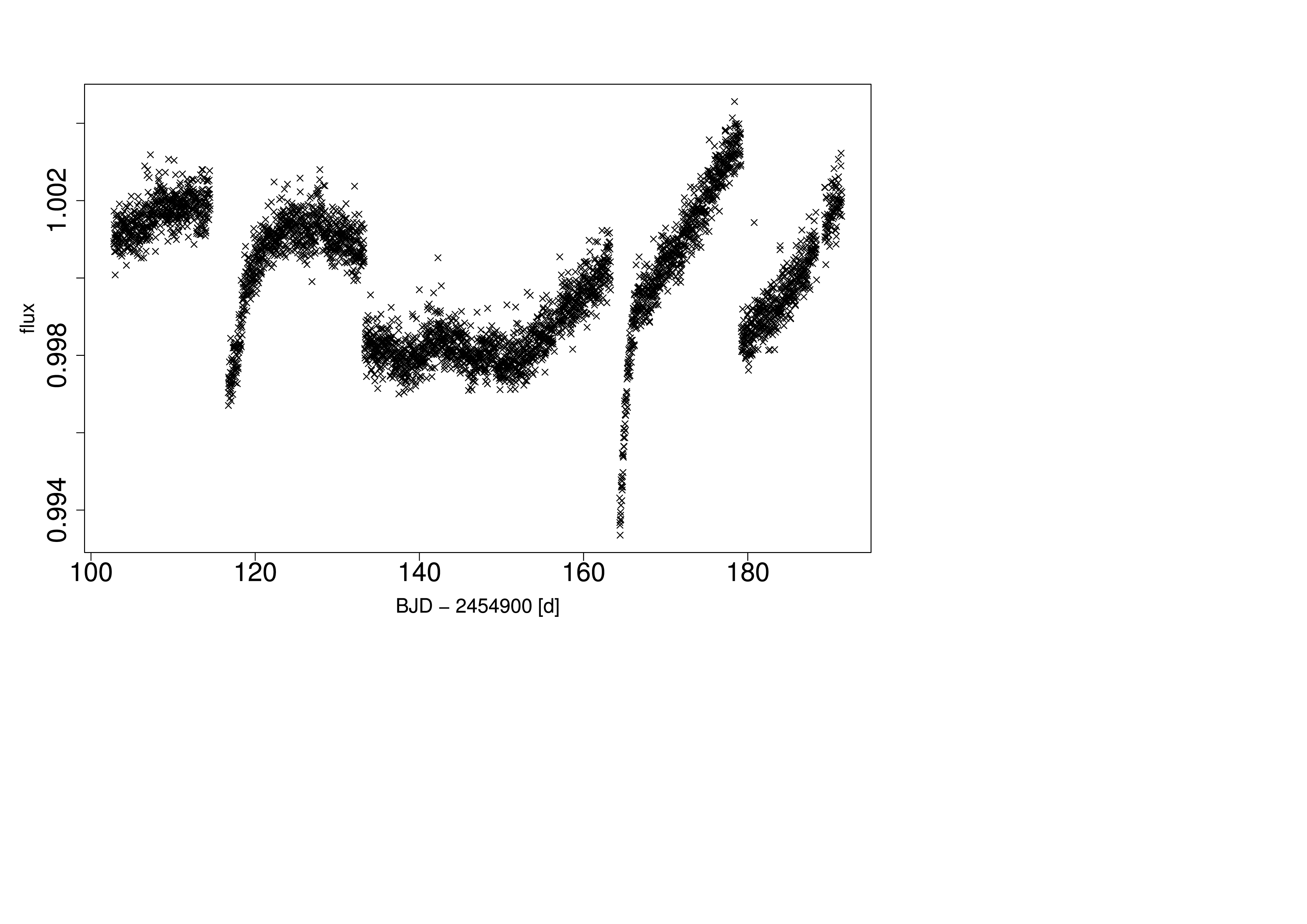}
\includegraphics[angle=270,scale=0.19]{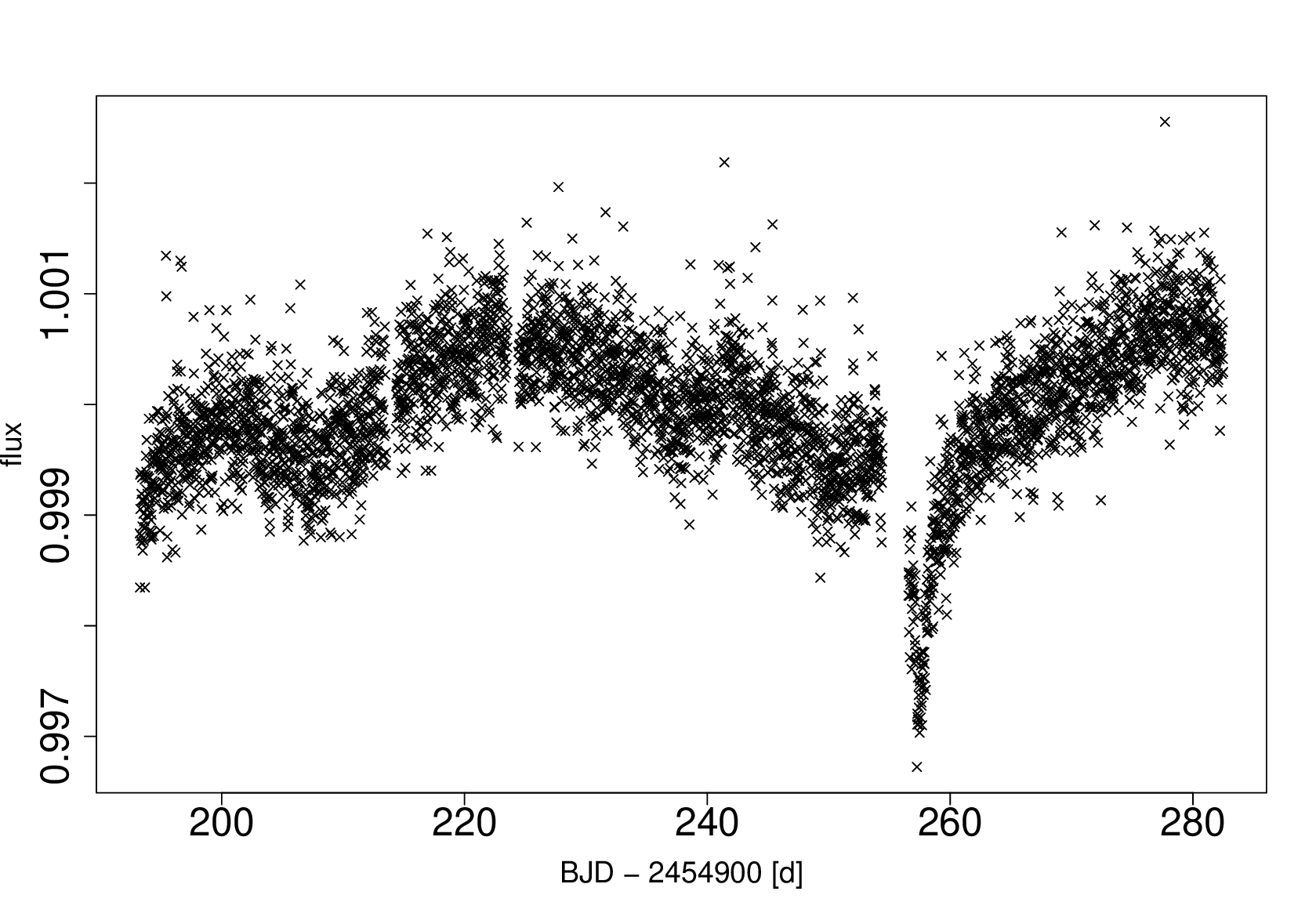}
 \caption{Quarter light curves of KIC07133671 for quarter 0-2 of public Kepler data. The vertical line denotes the position of one detected superflare (see Tab. \ref{properties}). {\bf Most of the light curve variations can be explained by a set of ``cotrending basis vectors'' \citep{smith}, that were designed to desribe the most common trends for all Kepler Field of View stars, caused by the spacecraft operation.}}
  \label{curves_2}
\end{figure*}
\section[]{Data Analysis and Results}
We used the Multimission Archive at Space Telescope Science Institute\footnote{http://archive.stsci.edu/kepler/} to retrieve the light curve and target pixel files (TPF) for each quarter. The light curve files contain exposure times, aperture photometry fluxes (SAP) with photometric errors and fluxes corrected for pointing errors, focus changes and thermal effects (PDC-SAP) \citep{jenkins}. In the TPFs, pixel data are presented as a time-series of photometrically calibrated images, while the pixels are assigned either to the optimal aperture or a halo around it \citep{kinemuchi}.

\subsection{Origin of photometric variability}

Fig. \ref{curves_1} and Fig. \ref{curves_2} show the PDC-SAP fluxes of KIC10524994 and KIC07133671 for each quarter. While there are 16 quarters available for KIC10524994, only 3 quarters could be gathered from the archive in the case of KIC07133671.
 
\begin{figure}
\includegraphics[width=1\columnwidth]{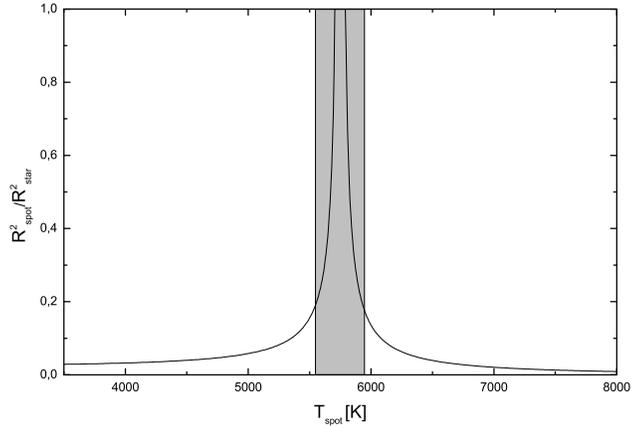} 
 %\vspace{pt}
 \caption{\bf Normalized spot size as a function of spot temperature for a central circular spot for KIC10524994. Spots across the solid line would induce brightness variations of the star equal to the $4\sigma$ range of the light curve variation due to their movement with the rotation of the star. We modelled the projected disc of the star with a linear limb darkening law \citep{claret} while the spots area was weighted with $(\text{T}_{\text{spot}}/\text{T}_{\text{star}})^4$. Within the $1\sigma$ range of T (grey zone), the spot size decreases to less than $20\%$ of the stellar hemisphere.}
  \label{spots}
\end{figure}

If we are visually looking at the quarterly light curves of KIC 10524994 (Fig. \ref{curves_1}), most of them show quasi-periodic brightness variations with an irregular modulation in amplitude of up to $40\:$mmag. A simple estimation shows that the averaged amplitude of these variations can be explained by stellar spots that have a slightly different temperature than the photosphere (Fig. \ref{spots}). If one considers similar spot temperatures as those for the Sun, which means $\approx4500\:$K for umbrae \citep{chaisson}, a single spot would cover about 4$\%$ of the visible hemisphere, {\bf which is a realistic scenario}. In contrast to binary eclipses with an expected constant amplitude, stellar spots can be variable regarding their spot size and temperature, while new spots can be generated at other longitudes and latitudes, hence one can explain the observed phase jumps (due to differential rotation) and amplitude shifts.

Although the hypothesis of binary-eclipses can not be ruled out for sure, we assume that the frequency of the most dominant light curve variation corresponds to the rotational period of the star. We used the Lomb-Scargle method (LSP) to calculate the spectral content of the signal. Rotational periods were estimated from the mean of a Gaussian shape, which was fitted to the highest peak of the power spectrum. We then compared the results of LSP with the results from computing the ``string length" \citep{dworetsky}. We normalized the magnitudes ($m_{i}$) and transformed all times into phases ($\phi_{i}$) that correspond to a particular period. We used the period for which the sum of the lengths of line segments between neighbouring points ($m_{i}$, $\phi_{i}$) is minimized as the true period. 

\begin{figure}
\includegraphics[width=1\columnwidth]{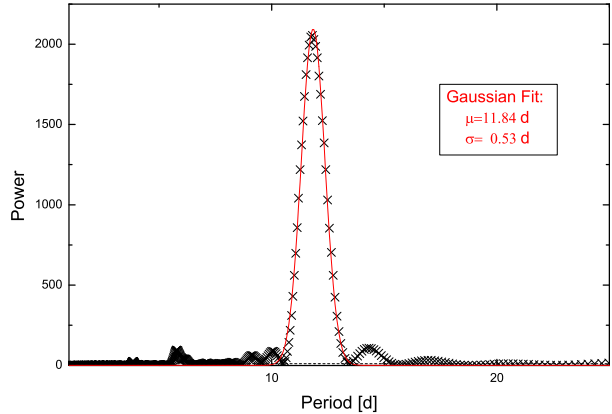} 
 %\vspace{pt}
 \caption{Lomb-Scargle-Periodogram of quarter 8 of KIC10524994. A period of $P_{rot}=(11.84\pm0.53)\:$d could be estimated from a Gaussian fit of the highest peak. Additionally to this main peak, one harmonic of that period at about $5.98\:$d is detected. For a better resolution, only the period range up to $25\:$d is shown, but was computed up to $50\:$d.}
  \label{fourier}
\end{figure}

From the Lomb-Scargle-Periodogram we got a period of $P_{rot}=(11.84\pm0.53)\:$d for KIC10524994 (Fig. \ref{fourier}), which is consistent with the value of \citet{maehara}.  

\begin{figure}
\includegraphics[angle=270, width=1\columnwidth]{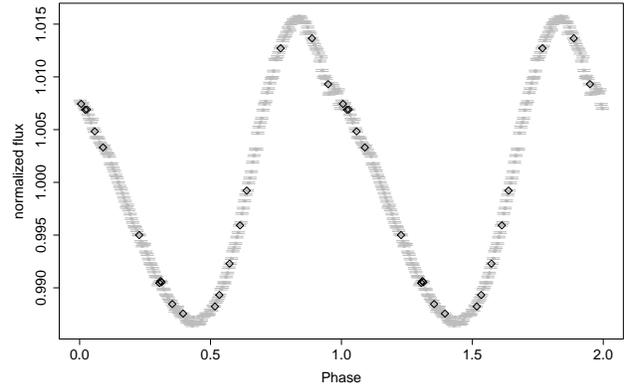} 
 %\vspace{pt}
 \caption{Phase folded and binned light curve of quarter 8 of KIC10524994 ($P_{rot}=11.87\:$d). The amplitude of the light curve can be explained by stellar spots, which are stable over some periods. {\bf Each data point represents the weighted mean of all data points within the corresponding phase bin, while the size of the error bars corresponds to their uncertainty.} The diamonds denote the occurrence phases of all detected superflares (see Tab. \ref{properties}). The flares are well distributed over the entire phase range.}
  \label{kurvenfaltung}
\end{figure}

We computed the string length for a spectral range between  $1\:$d and $50\:$d with a step rate of $2\ast10^{-5}\:\text{d}^{-1}$ in the frequency domain. Independent of the Lomb-Scargle method, a rotational period of $P_{rot}=(11.87\pm 0.07)\:$d could be determined, which is consistent with LSP. 
Fig. \ref{kurvenfaltung} shows the phase folded and binned light curve of KIC10524994 for quarter 8, which contains of the most stable amplitude.

From the light curve shape in Fig. \ref{kurvenfaltung} one can rule out some hypotheses: If the shape were be caused by eclipses of a binary, the disappearance of phases with constant flux would categorize the companion as a giant, hence it would dominate the spectral class of the system. An eclipsing binary in the background can be responsible for the amplitude, but it can not explain the brightness modulation in the different quarters and it is less realistic that such an amplitude is created by star spots of a single background star. The sinusoidal like light curve shape supports the hypothesis of large star spots on the stellar surface of KIC10524994 but nevertheless, the possibility of an unresolved binary or a combination of both eclipses and stellar spots remains. 

\begin{figure}
\includegraphics[angle=270, width=1\columnwidth]{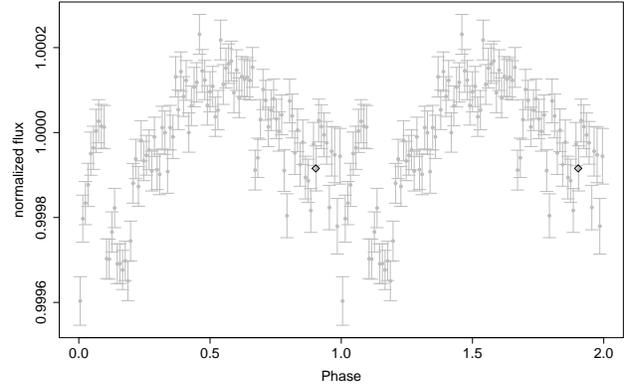} 
 %\vspace{pt}
 \caption{Phase folded and binned light curve of quarter 2 of KIC07133671 ($P=15.19\:$d). {\bf Each data point represents the weighted mean of all data points within the corresponding phase bin, while the size of the error bars corresponds to their uncertainty.} The diamond represent the occurrence phase of one detected superflare.}
  \label{kurvenfaltung2}
\end{figure}

{\bf The basic counting error for KIC10524994 is about $0.04\:$\% of the mean flux level ($0.03\:$\% for KIC07133671). Binned data points in Fig.  \ref{kurvenfaltung} and \ref{kurvenfaltung2} represent the weighted means for each set of data points that fall within a certain phase bin and the size of the error bars corresponds to their uncertainty. Since stellar spots can evolve rapidly during an observational quarter, data points of slightly different magnitudes and phase ranges might be combined here}. The occurrence phases of the detected superflares are well distributed over the entire phase range, as illustrated in Fig. \ref{kurvenfaltung} and Fig. \ref{kurvenfaltung2} by diamonds.

{\bf The quarter light curves of KIC07133671 are not quasi-periodic and most of the artefacts, plotted in Fig. \ref{curves_2}, can be explained by ``cotrending basis vectors'' \citep{smith}, that were designed to describe the most common trends of all Kepler targets.} Hence the light curve variations are not intrinsic features of the target. Nevertheless, we detected a period of $P=(15.19\pm0.85)\:$d with the ``string length" method \citep{dworetsky}, which is consistent with the rotational period found by \citet{shiba}. From the light curve shape (Fig. \ref{kurvenfaltung2}) one can not distinguish between an eclipsing binary or a star spot modulation. The brightness variation is up to $0.6\:$mmag. 
\subsection{Age determination from rotational periods} 
We used rotational periods to roughly estimate the ages of the stars. The upper limit was determined by a comparison with evolutionary models of the angular momentum as described in \cite{bouvier}. The main assumptions of this model are solid body rotation and interaction of the stellar surface with the time dependent appearance of a circumstellar disk \citep{bouvier}. A rotational period of $11.87\:$d (for solar radius) for KIC10524994 is consistent with a surface angular velocity of $\Omega \approx\ 2\Omega_{\odot}$. According to Fig. 4 in \citet{bouvier} we estimate an age of $\approx\:1.0^{+0.6}_{-0.4}\:$Gyr for KIC10524994, so that KIC10524994 might be less than one order of magnitude younger than the Sun. 

For KIC07133671 we do not estimate the age since it is very unlikely that the determined period is the rotation period of the star. A lower limit was estimated from proper motion measurements \citep{roeser} for both stars (see therefore Sect. 2.5).
\subsection{Flaring rate and cyclic behaviour}
\cite{maehara} analysed calibrated long-cadence Kepler flux time series of the first 3 quarters. The distribution of brightness changes between all pairs of 2 consecutive points were calculated to define a statistical threshold of a superflare event. The typical value of this threshold is about $0.1\:\%$ of the brightness of the stars in the Kepler filter response. In this work the analysis was extended to all available and public data of the first 16 quarters. To search for superflares and additionally smaller flares, we first smoothed the light curves with a 6th Level Discrete Wavelet Transformation (DWT). This works as a high pass filter that eliminates all intrinsic brightness variations with periods $>\:32\:$h while leaving all short time variabilities unaffected. Since a typical flare lasts only a few hours, this does not influence the superflare analysis. Then a Bayesian Analysis for Change Point Problems \citep{hartigan} was applied to the smoothed light curves. This analysis presents the probabilities of a sudden variation of the brightness for each time stamp of a light curve. Data points with a high probability for a sudden change ($> 50\:\%$) were fitted with a mathematical expression that is equivalent to a polynomial rising part and an exponential decay
\begin{eqnarray}
F(t) & = & \begin{cases} F_{0}+B*(t-t_{1})^{\alpha} \ \ \ \ \ \ \ \ \ t_{1}\leq t<t_{2} \\
 F_{0}+A*e^{-(t-t_{2})/t_{d}} \ \ \ \ \ \ \ \ \ \ \ \ \ t\geq t_{2} \end{cases}
\end{eqnarray}
to identify all the flares. $F_{0}$ is the normalized flux, $A$ is the flare amplitude, $\alpha$ describes the power law of impulsive phase, $t_{1,2}$ denote the flare start and peak time, and $t_{d}$ is a decay constant that corresponds to the duration of the flare. These parameters were optimized with a Levenberg-Marquardt routine and a simple F-Test with a commonly used $95\:\%$-Confidence interval was applied to compare our result with a constant model. 

\begin{figure*}

\includegraphics[angle=270,scale=0.15]{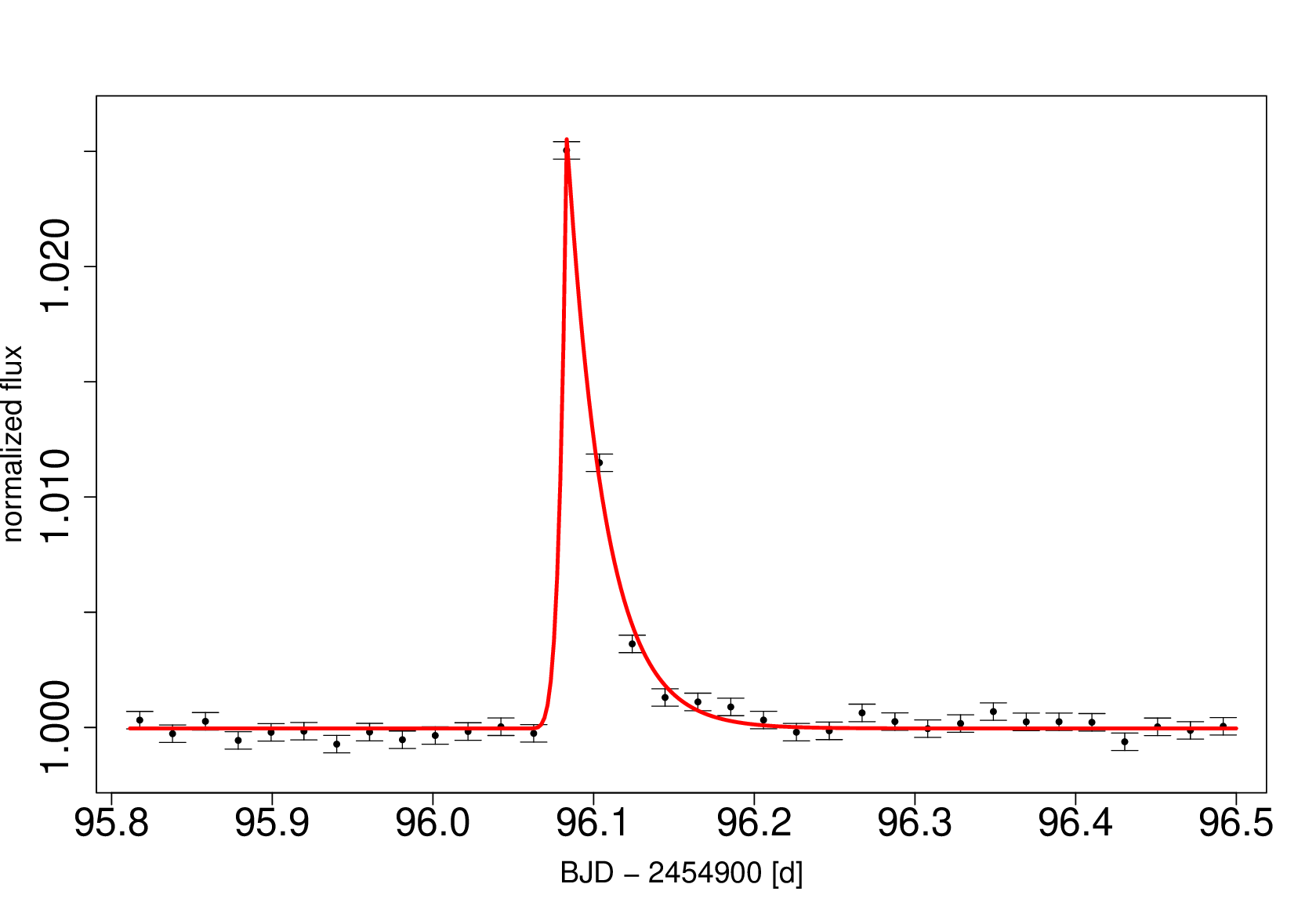}
\includegraphics[angle=270,scale=0.15]{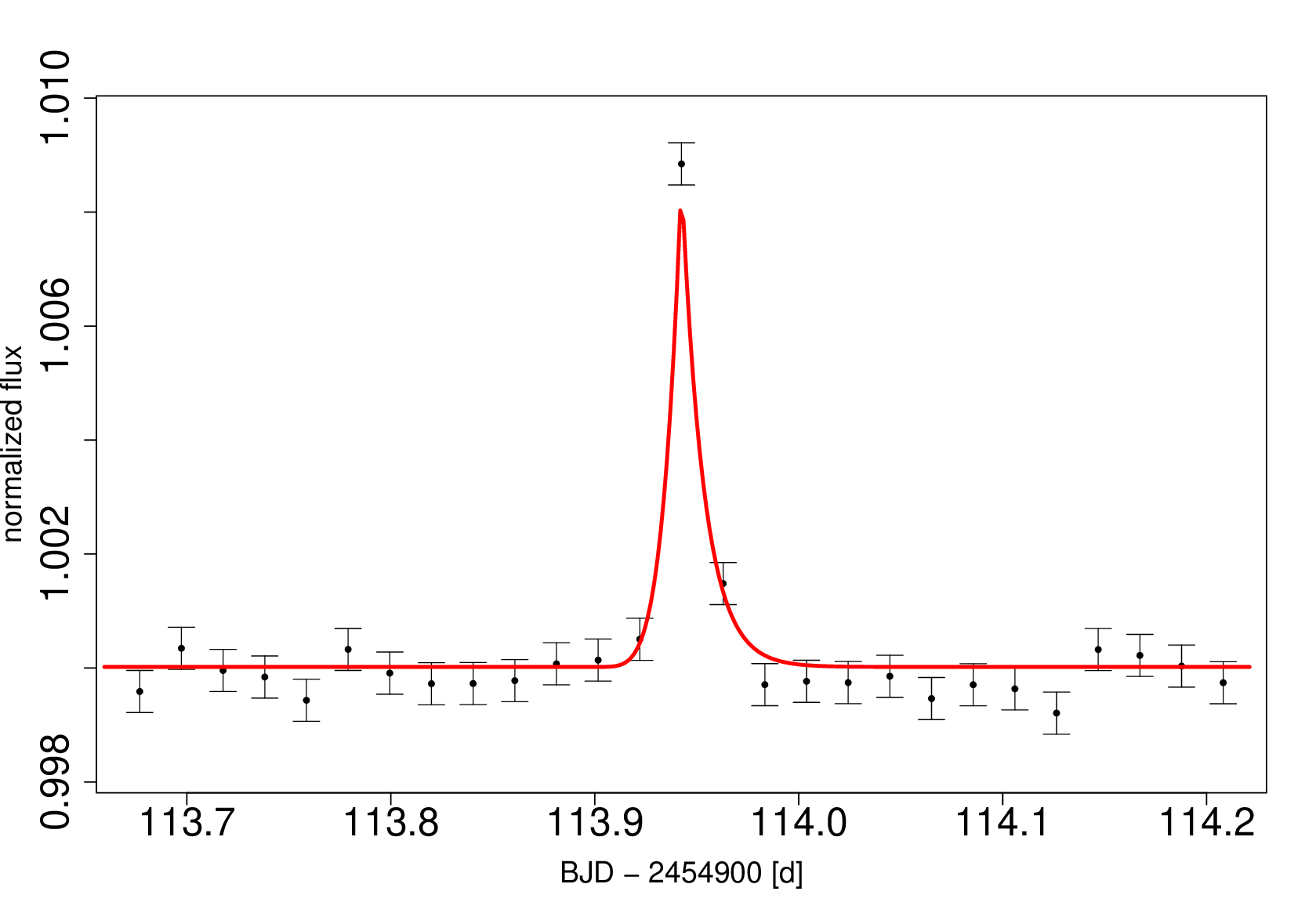}
\includegraphics[angle=270,scale=0.15]{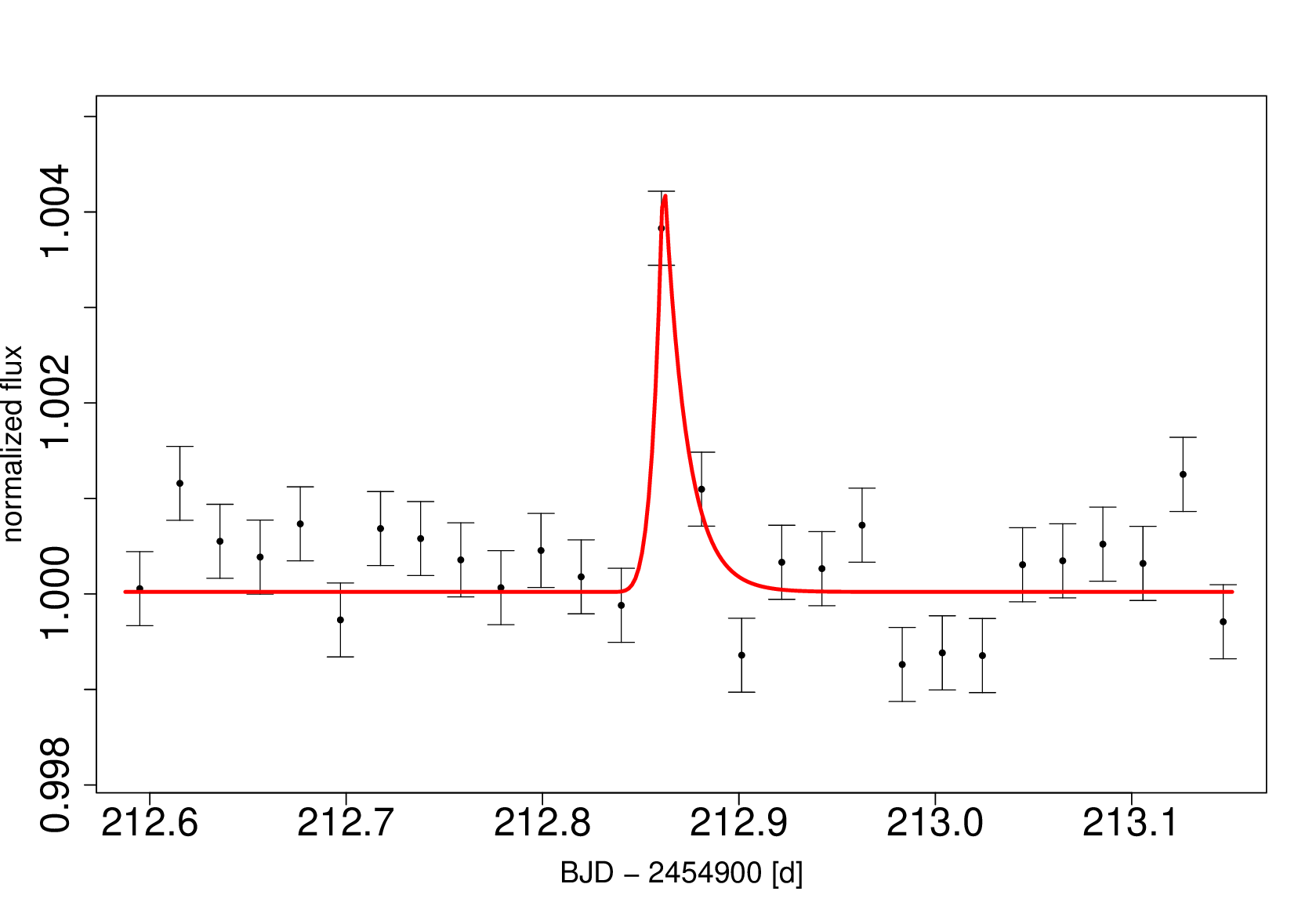}
\includegraphics[angle=270,scale=0.15]{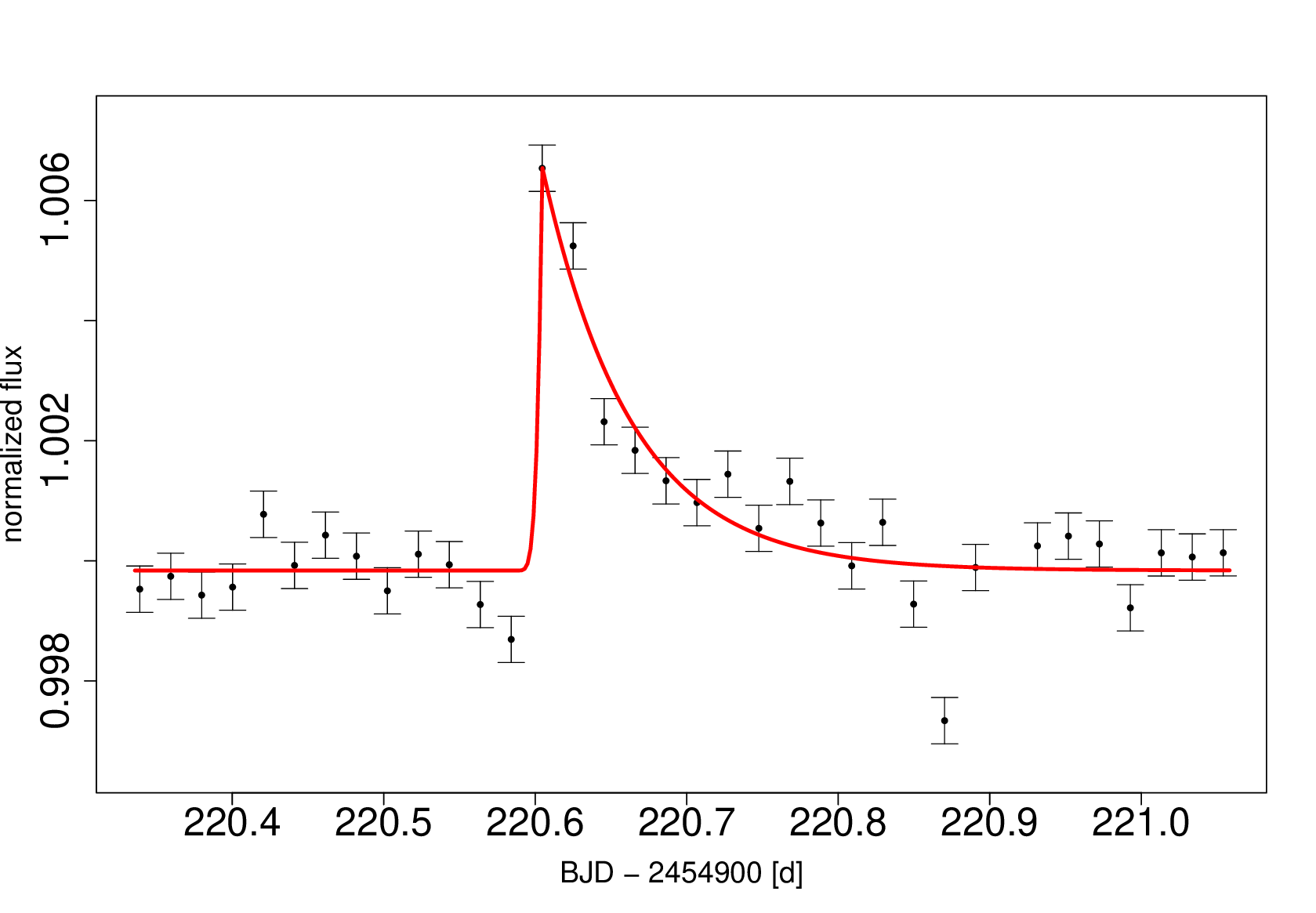}
\includegraphics[angle=270,scale=0.15]{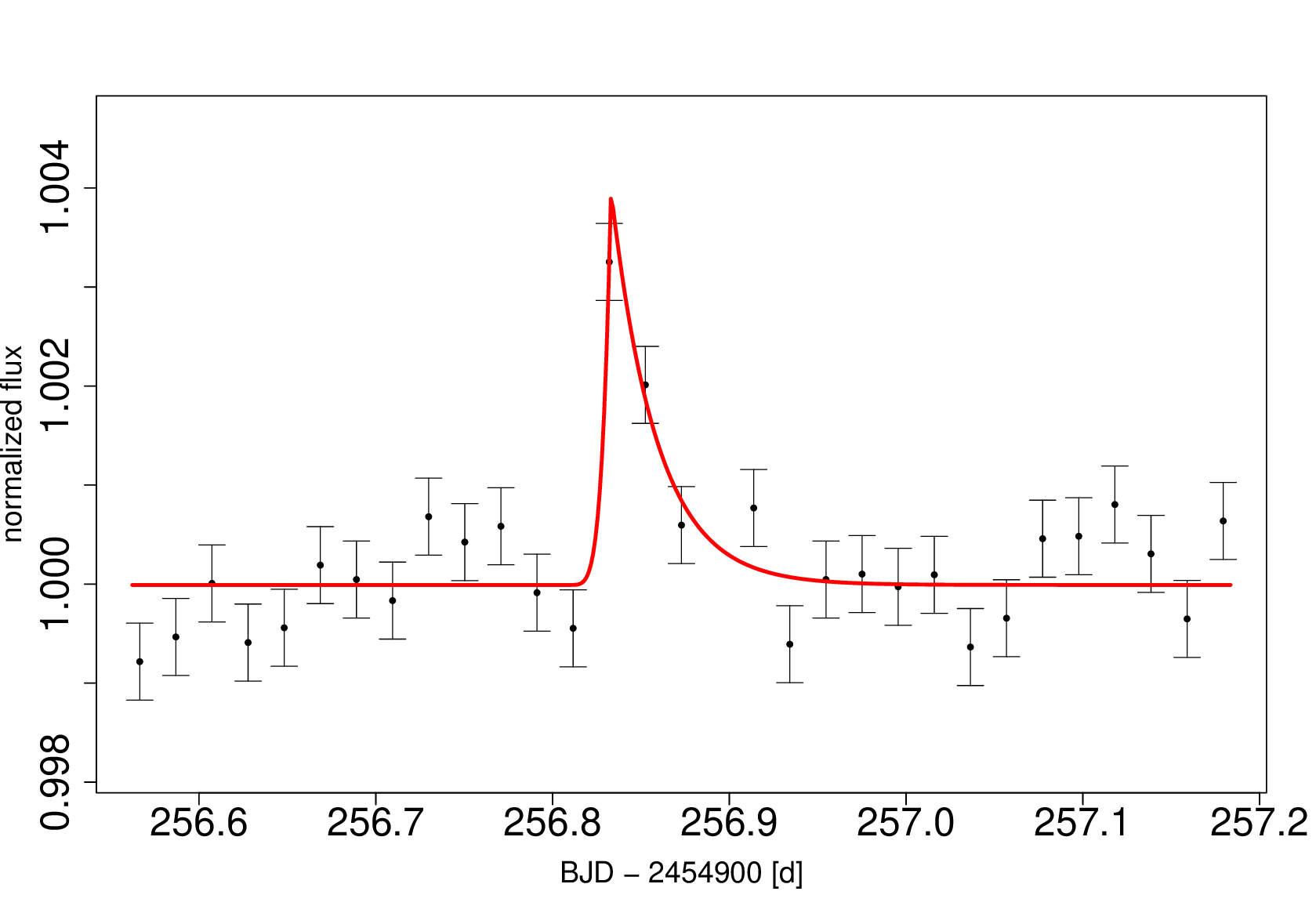}
\includegraphics[angle=270,scale=0.15]{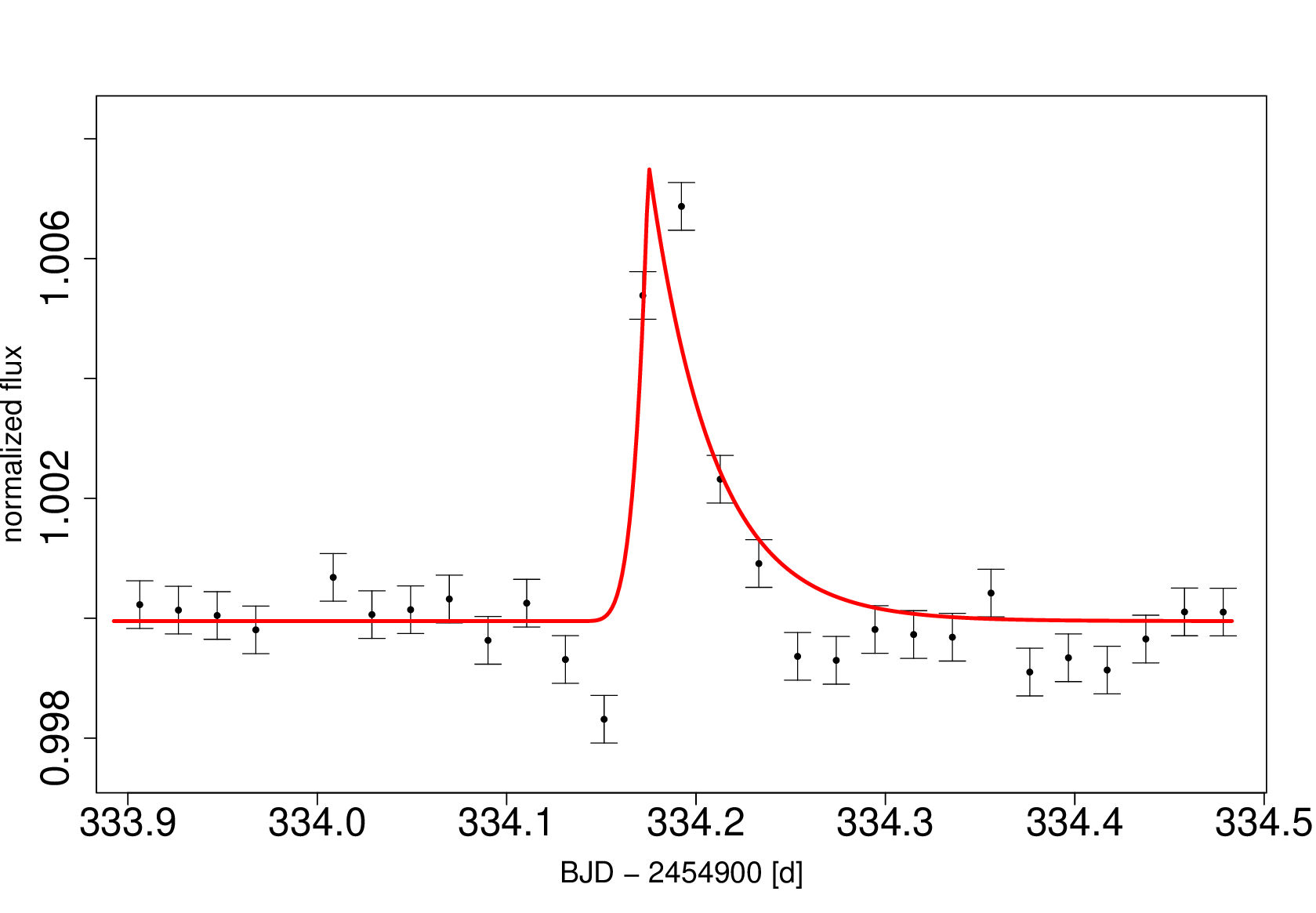}
\includegraphics[angle=270,scale=0.15]{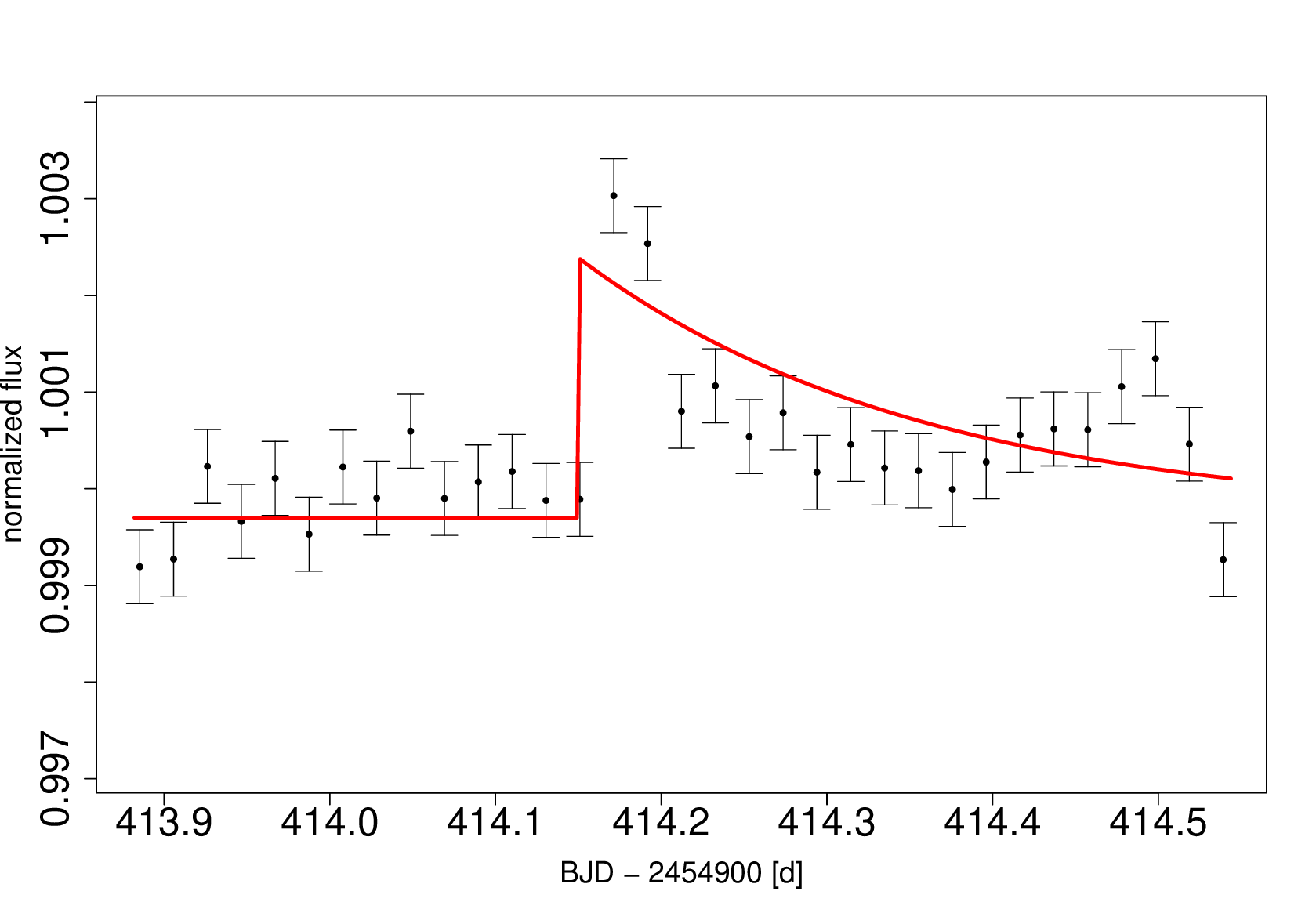}
\includegraphics[angle=270,scale=0.15]{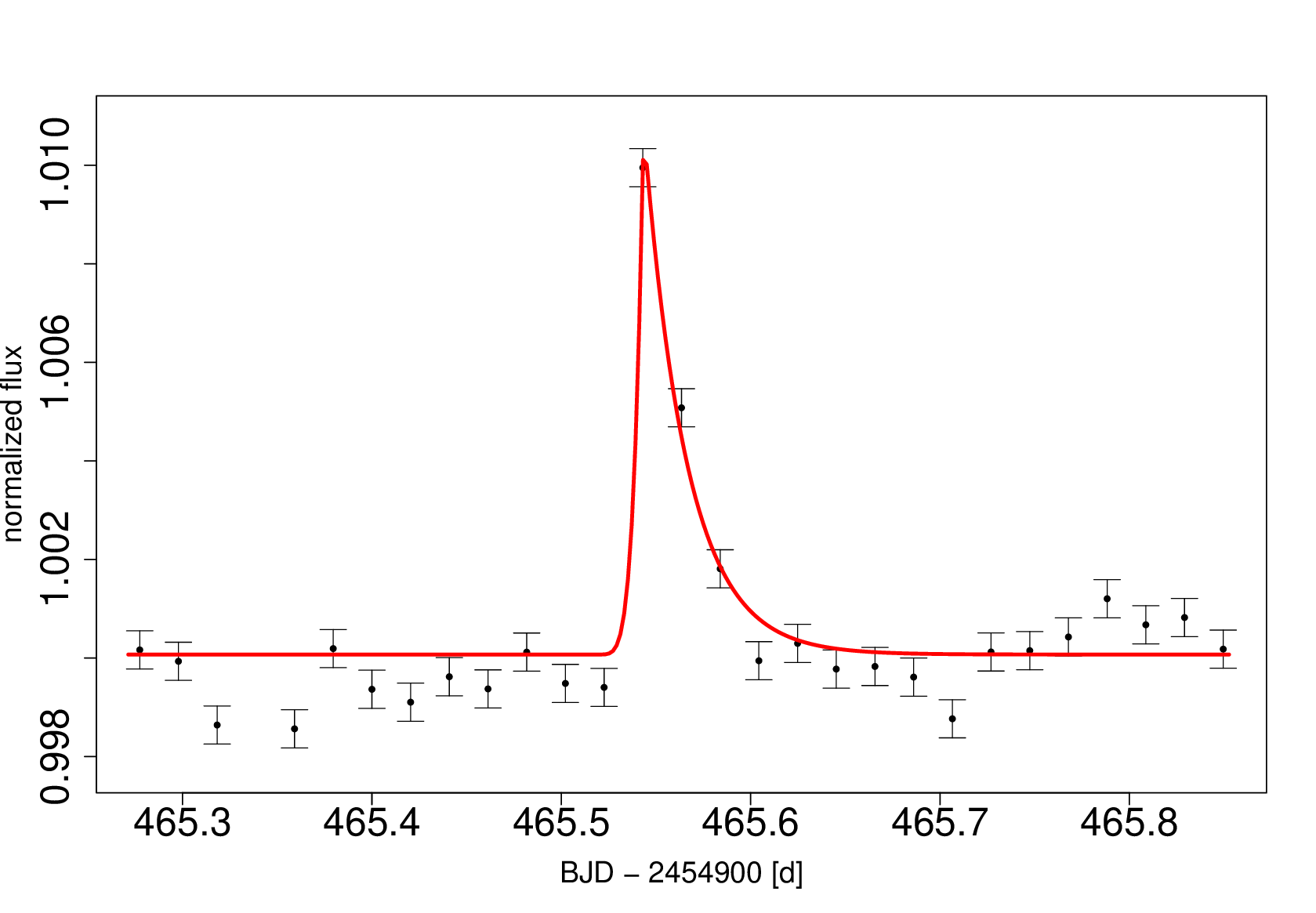}
\includegraphics[angle=270,scale=0.15]{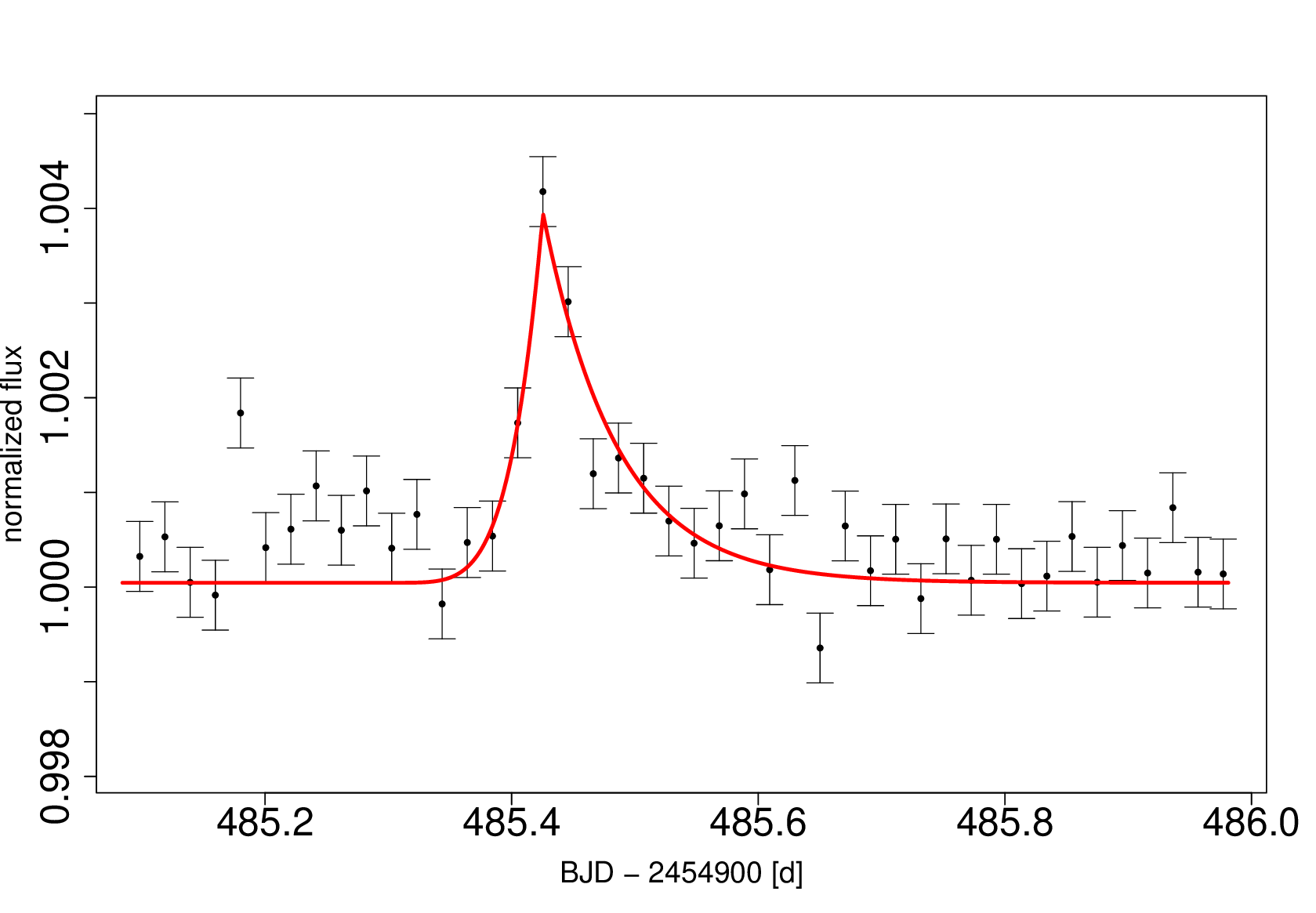}
\includegraphics[angle=270,scale=0.15]{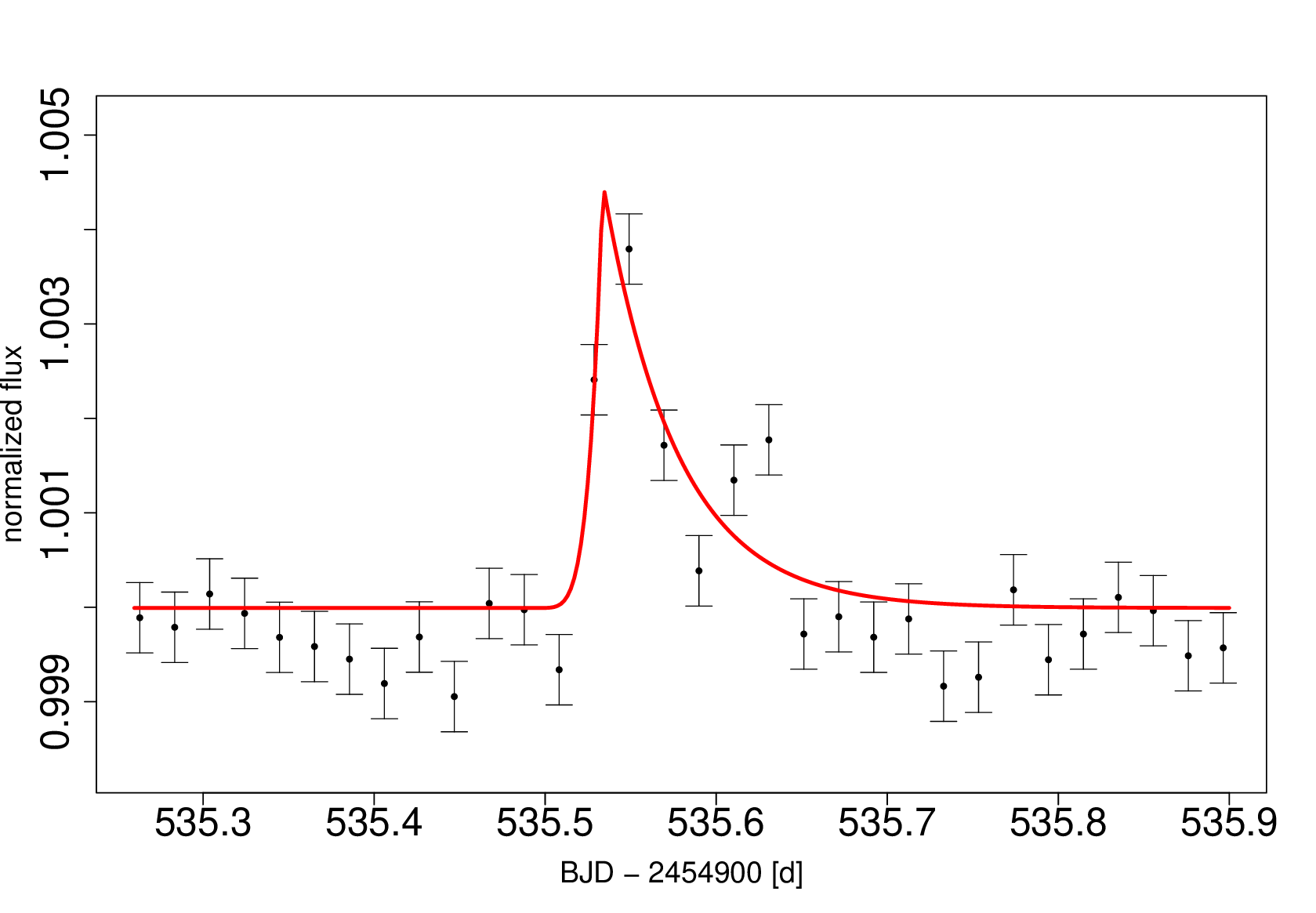}
\includegraphics[angle=270,scale=0.15]{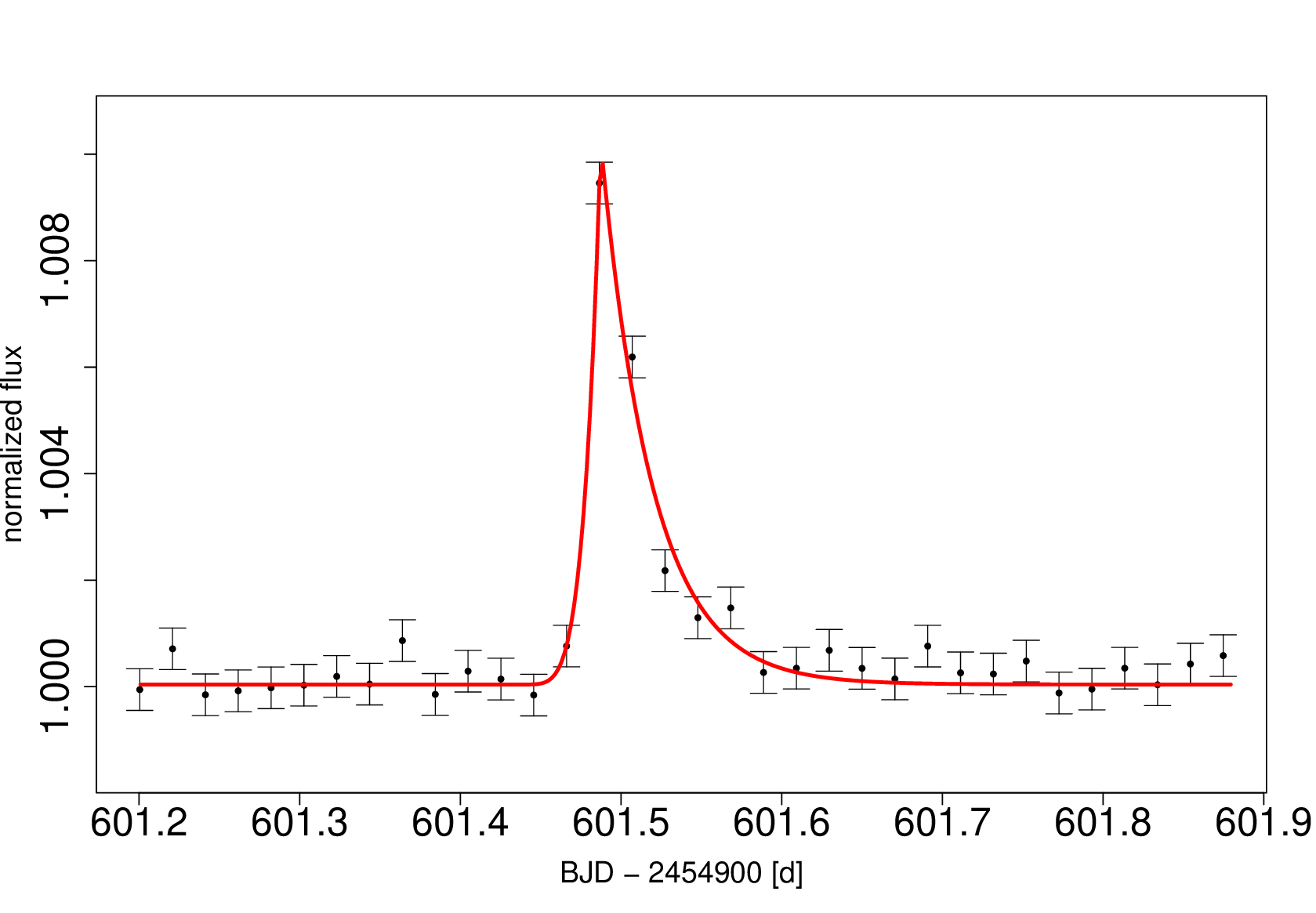}
\includegraphics[angle=270,scale=0.15]{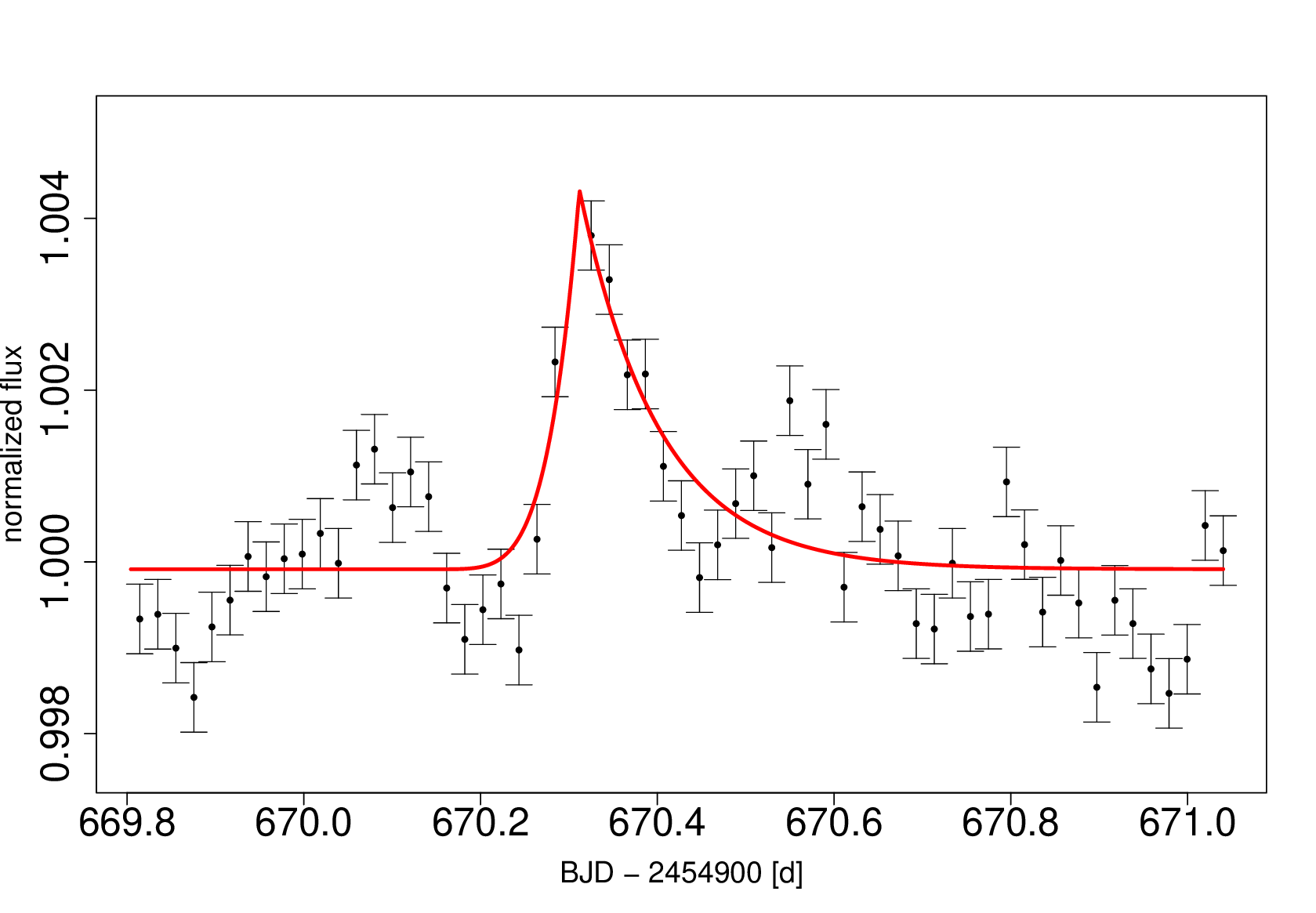}
\includegraphics[angle=270,scale=0.15]{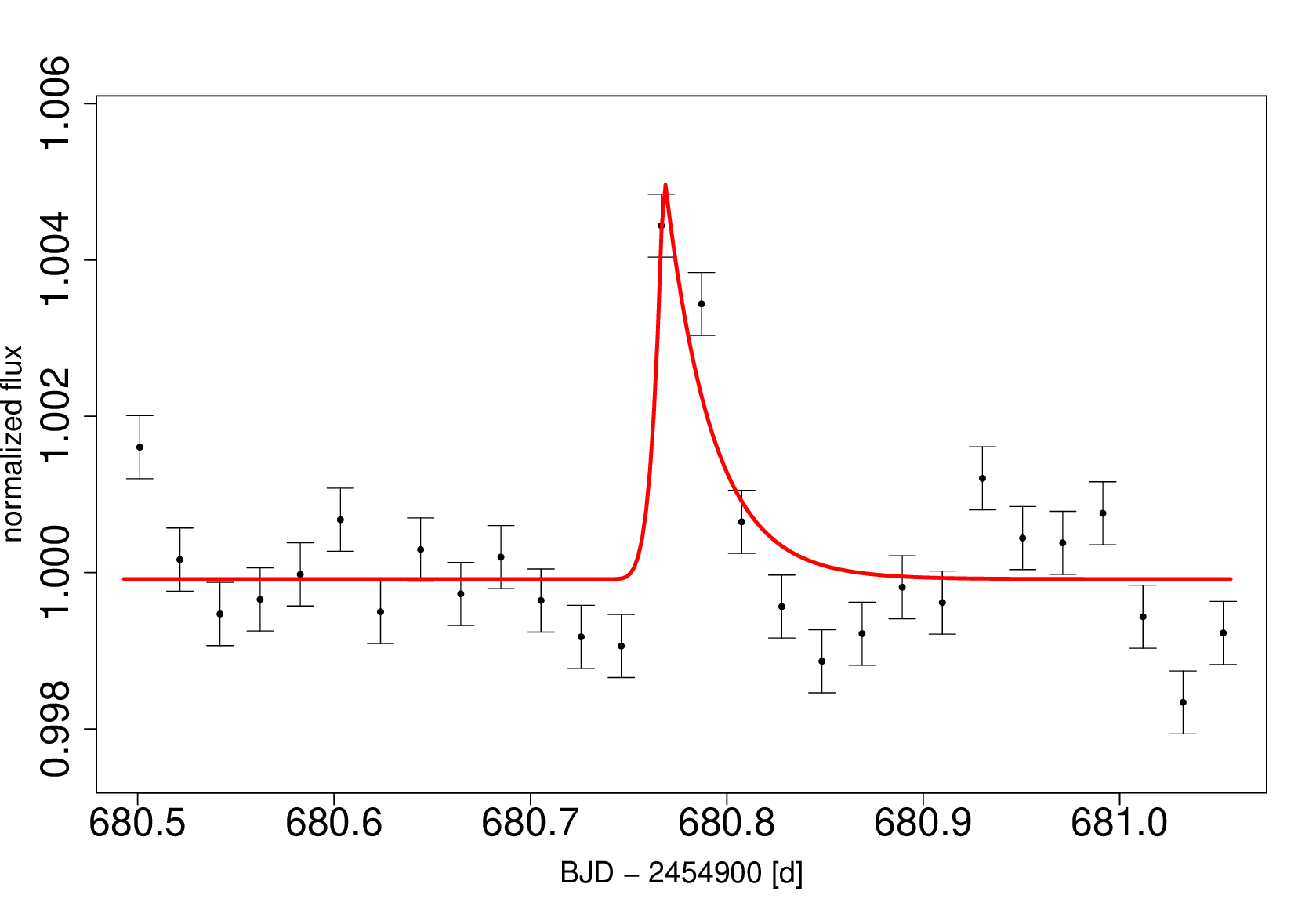}
\includegraphics[angle=270,scale=0.15]{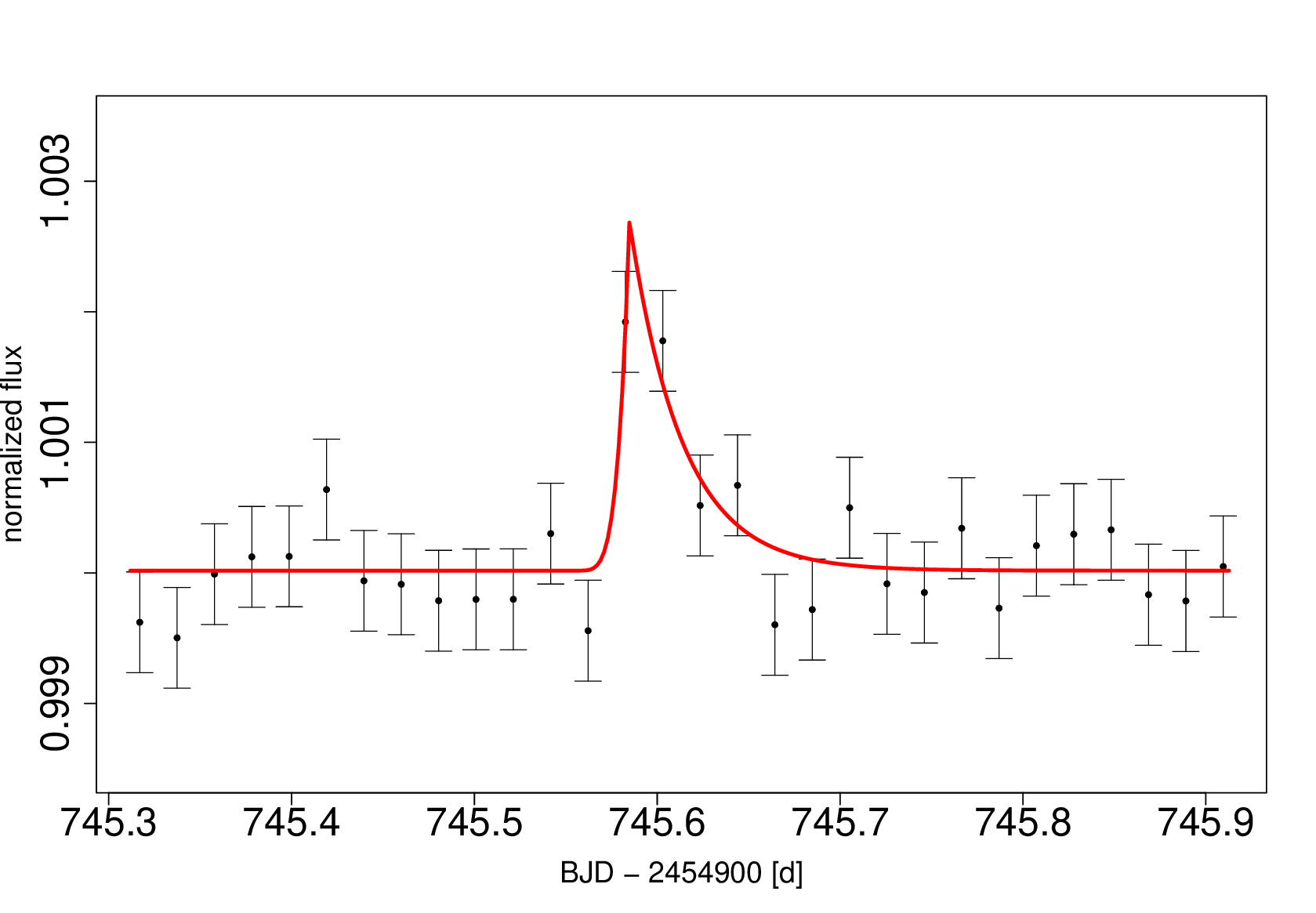}
\includegraphics[angle=270,scale=0.15]{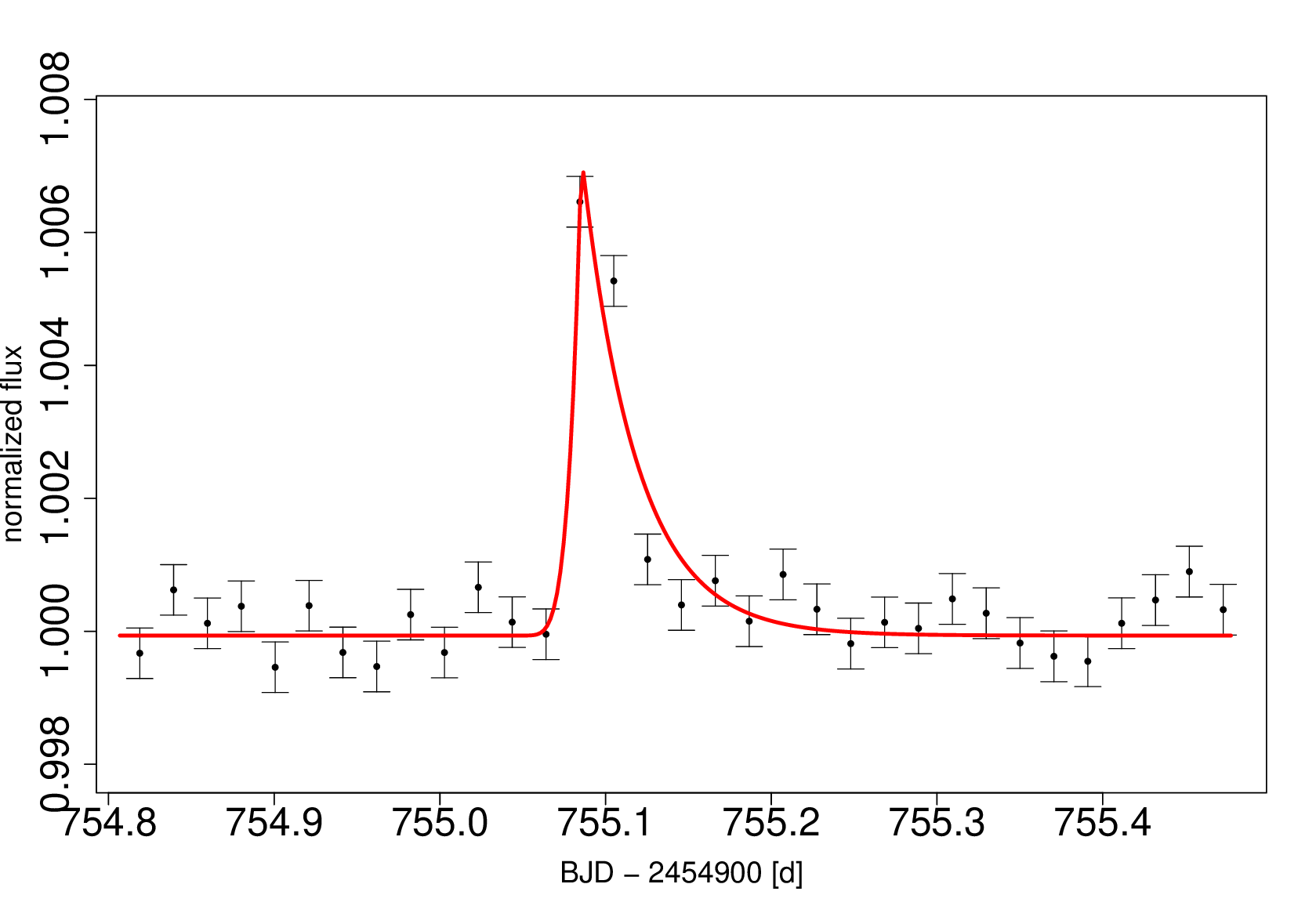}
\includegraphics[angle=270,scale=0.15]{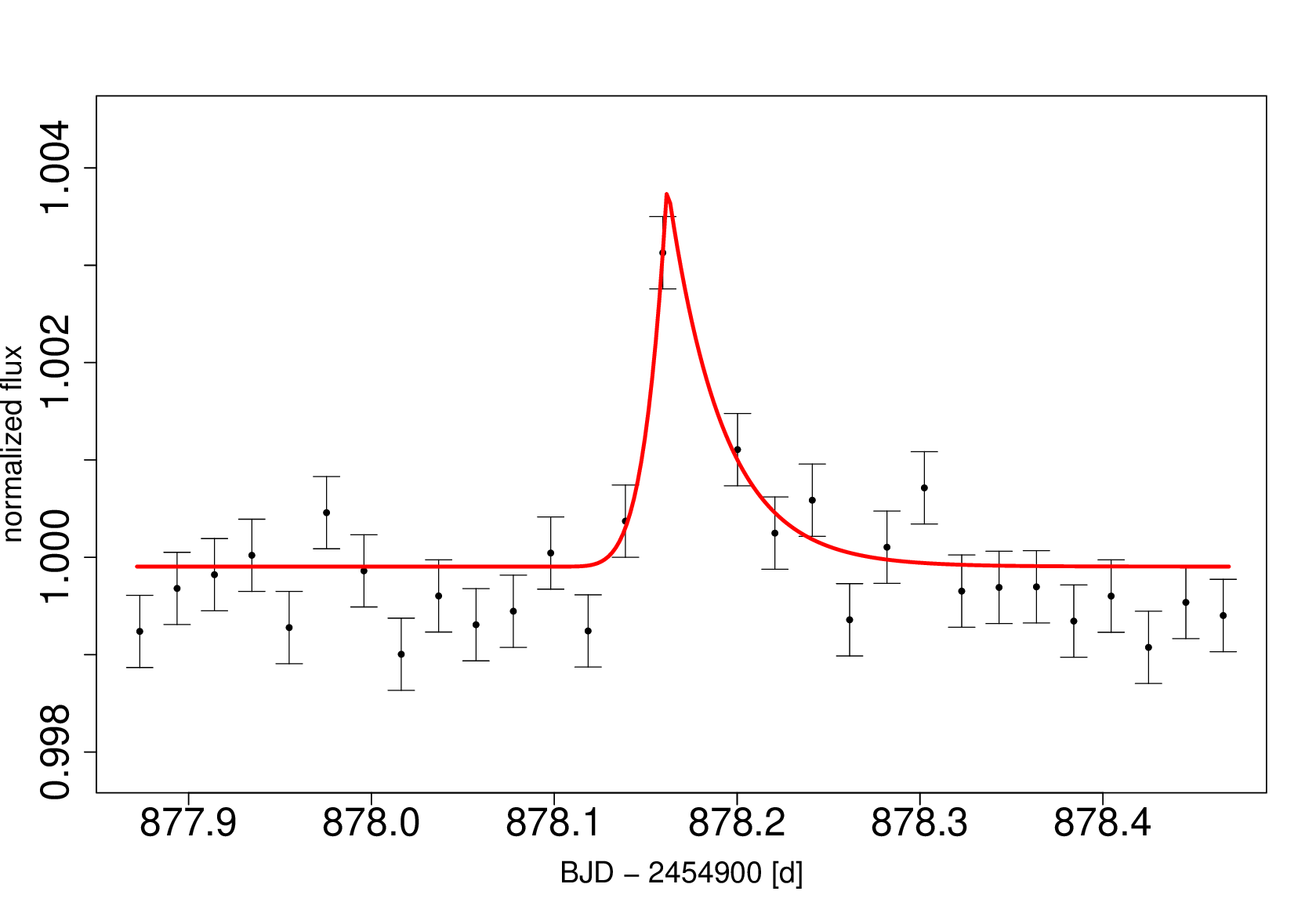}
\includegraphics[angle=270,scale=0.15]{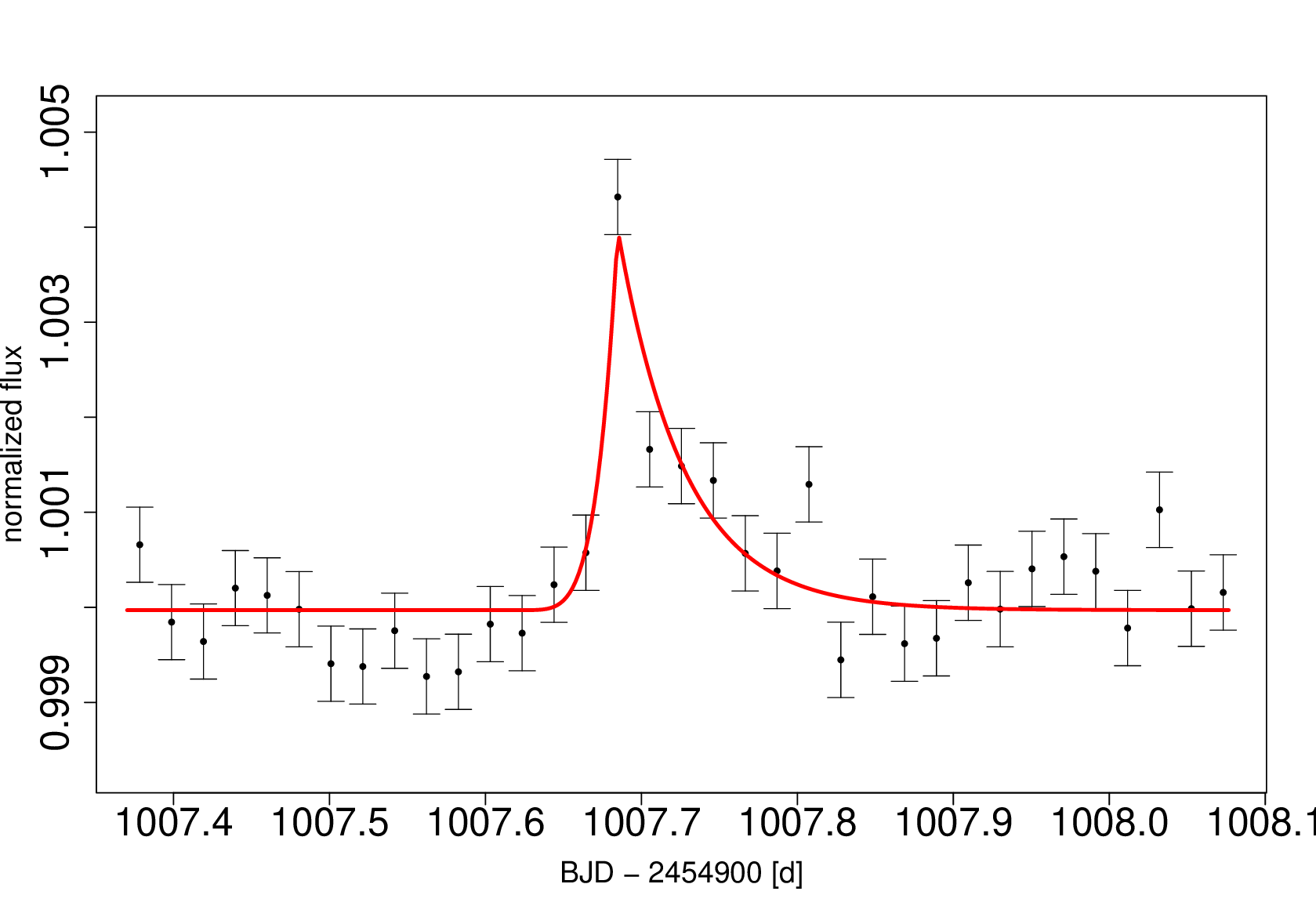}
\includegraphics[angle=270,scale=0.15]{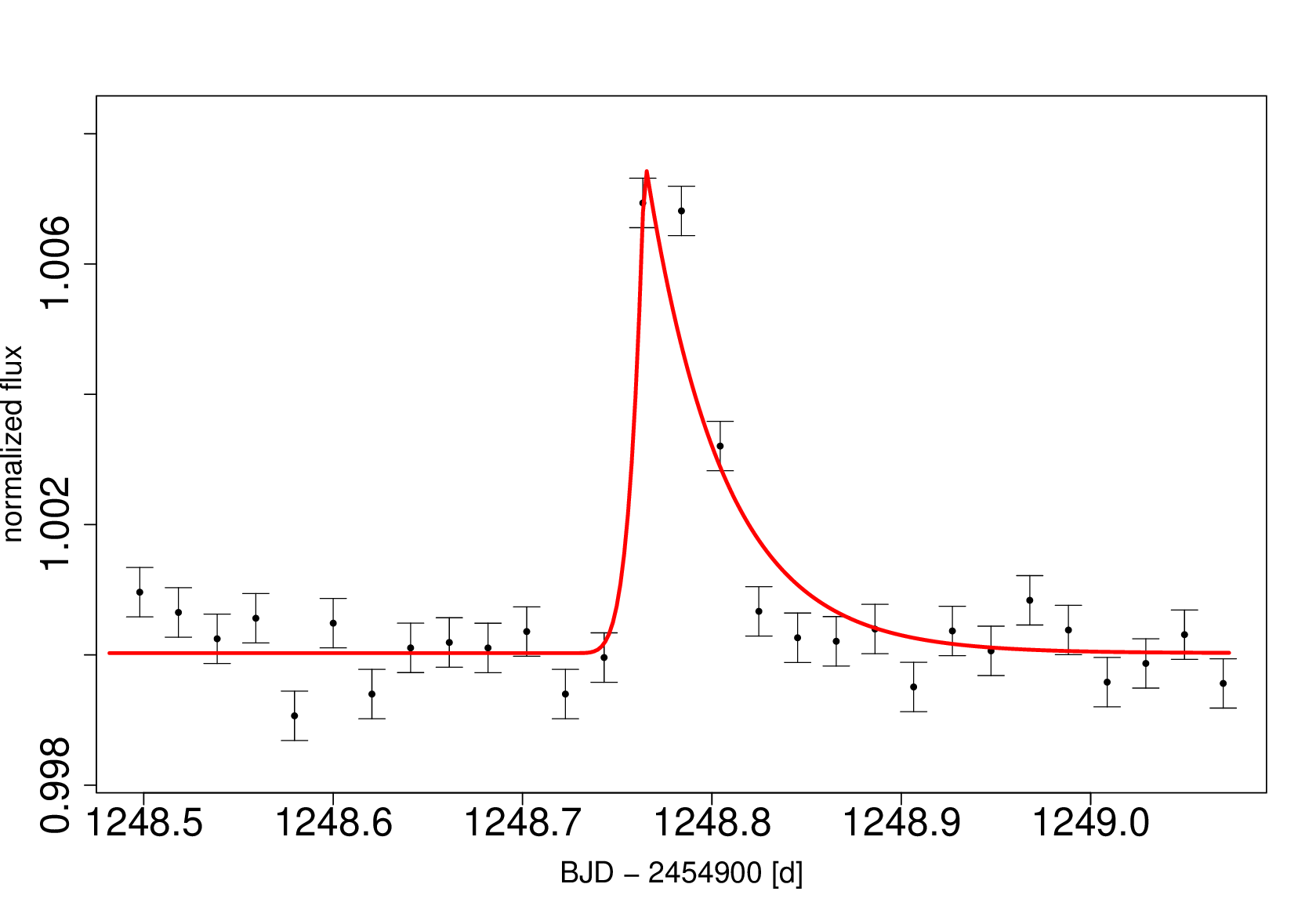}
\includegraphics[angle=270,scale=0.15]{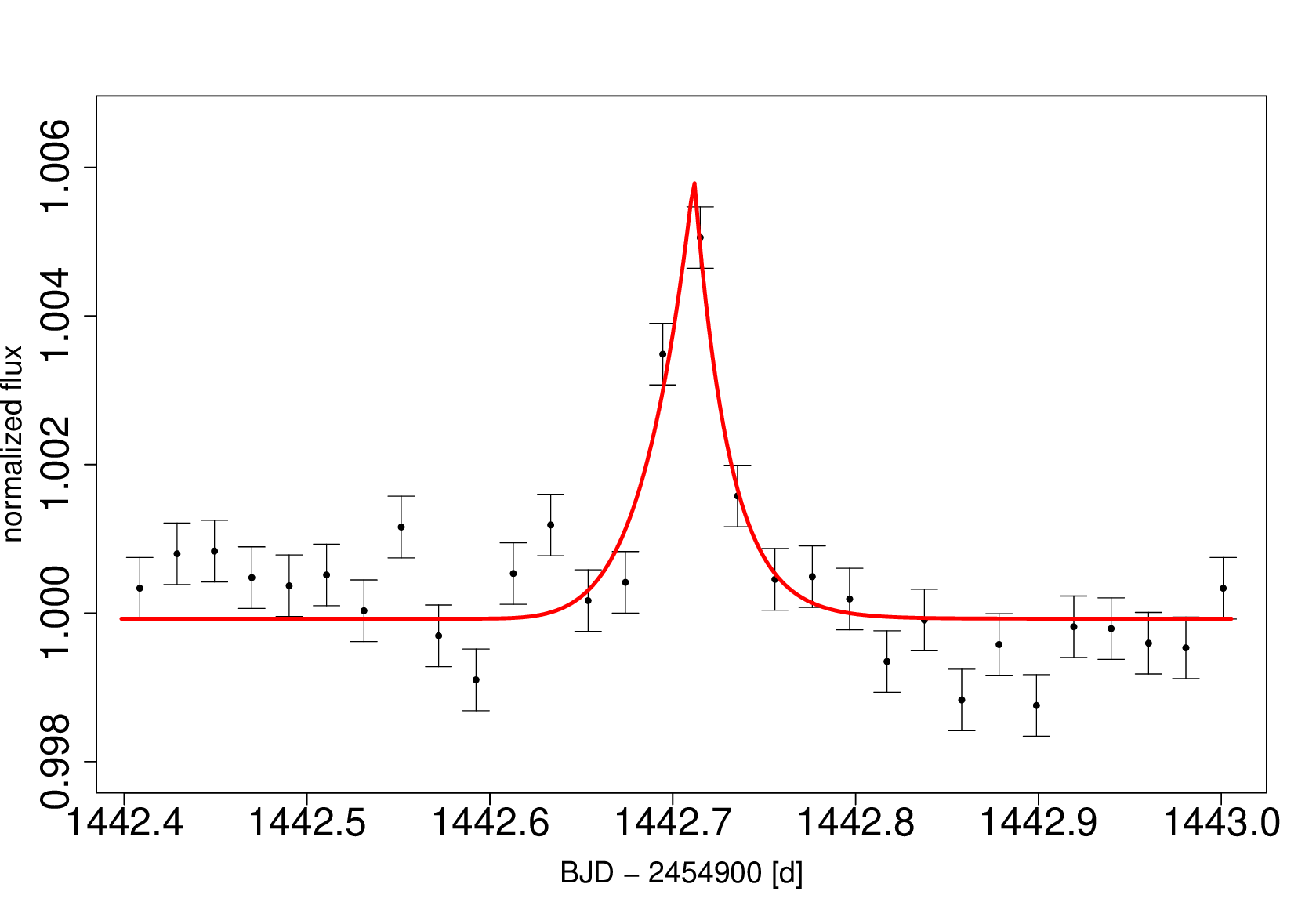}
\includegraphics[angle=270,scale=0.15]{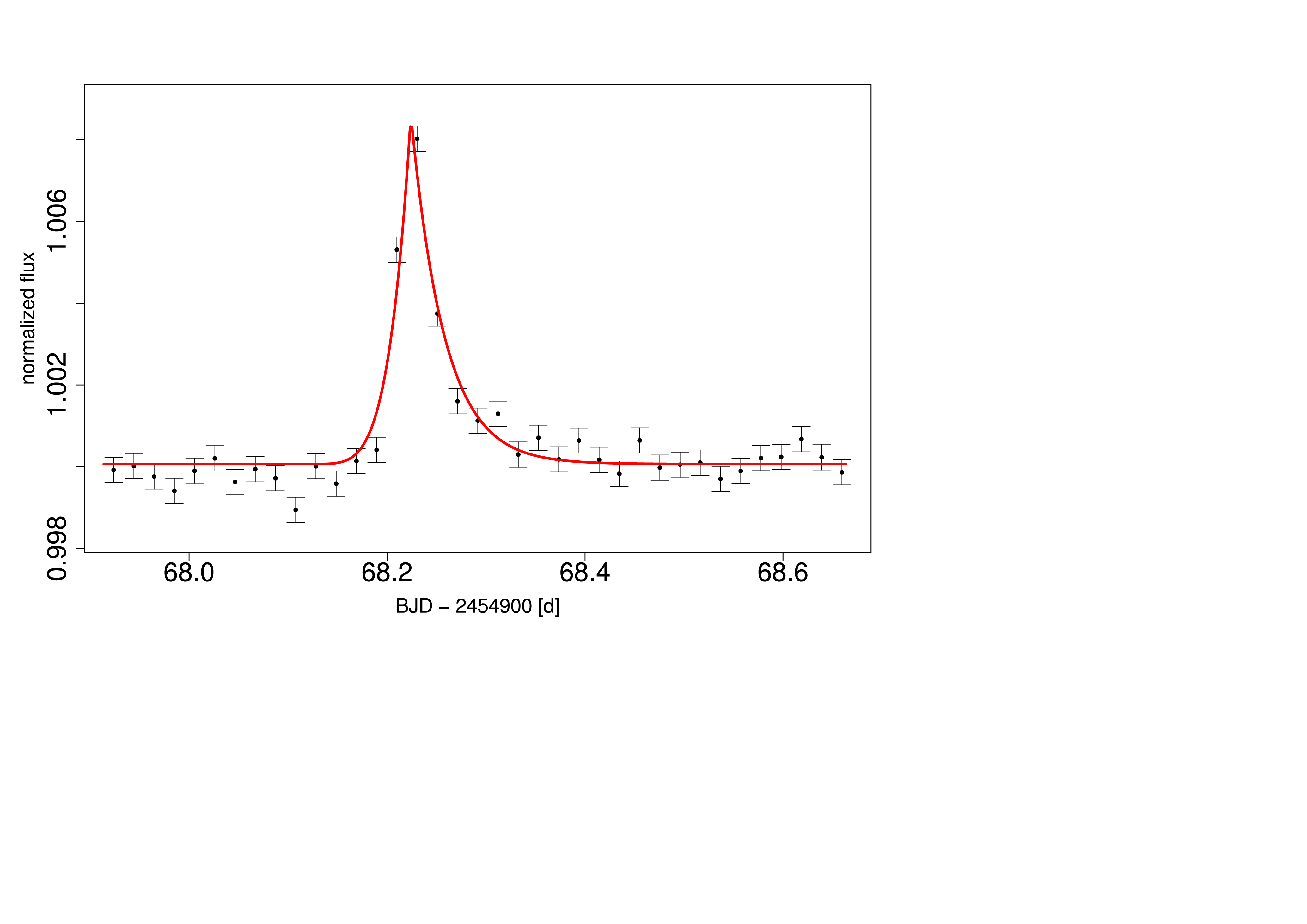}
 %\vspace{pt}
 \caption{Detected flares in the light curves of KIC07133671 (lower right) and KIC10524994 (all others, see also Tab. \ref{properties}). The normalized fluxes are plotted over time. All the flares were fitted with a Levenberg-Marquardt optimization routine. Due to the time resolution of $29.4\:$min between consecutive data points, only flares with a duration longer than about $30\:$min could be detected. It is therefore expected that the stars have many more flares on shorter time scales.}
  \label{flares}
\end{figure*}

We could detect 4 new superflares during the observing epoch \citet{maehara} investigated and 14 additional new superflares in the remaining quarters for KIC10524994 (in total 19 superflares) by performing a Change Point analysis \citep{hartigan} with 1000 iterations to consecutive segments of the light curves with lengths of $4\:$d. For KIC07133671 we could detect one superflare, the one observed by \citet{maehara}. Each single flare is shown in Fig. \ref{flares}.

The occurrence times of these flares were used to investigate any cyclic properties. It is possible that close binaries and additional ``hot jupiters" in eccentric orbits can cause superflares by strong magnetic interaction with their host stars \citep{ruben}. In this case one might expect a periodic behaviour of the flare frequency that could be higher during the passage of periastron and lower in the remaining orbit. We used the method of \citet{gregory} to compare a constant model with the hypothesis of a periodic intensity rate. From these calculations we can not confirm a close companion around KIC10524994, which might be biased by the low number of detected flares. Additionally we have not detected planetary transits among our targets.

\subsection{Flare properties}
\begin{table}
\centering
\caption{KIC Data for KIC10524994 and KIC07133671. All the quantities could be derived from $T_{eff}$ and $\log(g)$. Their uncertainties are estimated to be $\pm 200\:$K and $\pm 0.3\:$dex \citep{brown}.}
\begin{tabular}{ccc}\toprule
 & KIC10524994 & KIC07133671\\ 
 \midrule
$\textit{T}_{\textit{eff}}$ $[\text{K}]$    & $5747\pm200$ & $5657\pm200$\\
$\textit{R}_{\textit{star}}$ $[\text{R}_{\odot}$] & $0.971\pm0.335$ & $1.140\pm0.394$\\
$\textit{log(g)}$ $[\text{dex}]$ & $4.5\pm0.3$& $4.4\pm0.3$\\
$\textit{mag}_{\textit{kep}} $ & $15.34\pm0.03$ & $15.47\pm0.03$\\
$\textit{mag}_{\textit{K}}$ & $13.67\pm0.04$& $13.73\pm0.06$\\
$\textit{A}_{\textit{V}}$ $[\text{mag}]$ & 0.402& 0.581\\
\bottomrule
\end{tabular}
\label{KIC_data}
\end{table}
 
\begin{table}
\centering
\caption{List of all detected flares (Fig. \ref{flares}) with their occurrence time, bolometric luminosity, energy, and equivalent duration $d_{equ}$ for KIC10524994 and KIC07133671 (values for a $10000\:$K black body radiation). The flare energy was calculated by integrating the estimated luminosity over time from the start to the end of the flare. The end of the flare was determined by F-Test statistics.}
\begin{tabular}{crrrr}\toprule
\# & \multicolumn{1}{c}{MBJD\footnotemark{} [d]} & \multicolumn{1}{r}{$L_{bol}$ $[10^{32}\:$erg/s]} & \multicolumn{1}{r}{$E_{bol}$ $[10^{35}\:$erg]} & \multicolumn{1}{r}{$d_{equ}$ [s]}\\
\midrule
\multicolumn{5}{c}{KIC10524994}\\
\midrule
1 &96.06	&$1.05\pm	0.93$ &	$2.68\pm	2.38 $&	74.66   \\
2 &113.89&	$0.30\pm	0.26$ &$	0.88\pm	0.78$ &	24.45   \\
3 &212.80&	$0.12\pm	0.11$ &$	0.38\pm	0.34$ &	10.51  \\
4 &220.58&	$0.27\pm	0.24$ &$	1.34\pm	1.19$ &	37.44  \\
5 &256.81&	$0.19\pm	0.17$ &$	0.43\pm	0.39$ &	12.08    \\
6 &334.14&	$0.34\pm	0.30$ &$	1.04\pm	0.92$ &	28.96    \\
7 &414.15&  $0.16\pm    0.14$ &$	0.50\pm 0.44$ & 13.90	\\
8 &465.52&	$0.46\pm	0.41$ &$	1.09\pm	0.97$ &	30.54    \\
9& 485.33&	$0.15\pm	0.13$ &$	1.32\pm	1.18$ &	36.87   \\
10& 535.51&	$0.21\pm	0.19$ &$	0.75\pm	0.67$ &	20.94  \\
11& 601.45&	$0.44\pm	0.39$ &$	1.58\pm	1.40$ &	44.06  \\
12& 670.18& $0.16\pm 0.14$ & $1.73\pm1.54$ & 48.34\\
13& 680.74&	$0.25\pm	0.22$ &$	0.55\pm	0.49$ &	15.42   \\
14& 745.56&	$0.15\pm	0.13$ &$	0.35\pm	0.31$ &	9.83    \\
15& 755.06&	$0.33\pm	0.29$ &$	1.03\pm	0.92$ &	28.85  \\
16& 878.12&	$0.19\pm	0.17$ &$	0.57\pm	0.51$ &	16.03  \\
17& 1007.61&	$0.15\pm	0.13$ &$	0.85\pm	0.75$ &	23.61  \\
18& 1248.73&	$0.35\pm	0.31$ &$	1.21\pm	1.07$ &	33.67  \\
19&1442.65 & $0.28\pm0.25$& $0.88\pm0.78$& 24.60\\
\midrule
\multicolumn{5}{c}{KIC07133671}\\
\midrule
1 & 68.20 & $0.48\pm0.43$ & $2.20\pm2.00$ & 47.45\\
\bottomrule
\end{tabular}
\label{properties}
\end{table}
All detected flares and superflares with their fitting parameters could be used for further characterization. We were interested in the distribution of bolometric flare luminosities, energies and their influence on the waiting time between two consecutive events. If one observes only in one pass band, one can not determine the spectral content of flares. Moreover, observations on the Sun indicate that the spectral composition of radiated flare energy is a function of time, since the light curve shape of a flare is different when observing in different pass bands \citep{Benz}. We simplified the complex behaviour of a flare event to be an emission of a black body radiation, as suggested by \citet{kretzschmar} to estimate bolometric flare energies and luminosities. We assumed higher temperatures of the black bodies than the stellar photosphere and therefore calculated bolometric flare energies and luminosities representative for $6000\:$K, $10000\:$K and $20000\:$K. \footnotetext{MBJD = BJD - 2454900 [d]}Values in Tab. \ref{properties} are calculated for $10000\:$K. Values for $6000\:$K and $20000\:$K can be derived by scaling factors of $0.82$ and $4.79$, respectively. 

To calculate bolometric flare energies and luminosities, additional data can be obtained from Tab. \ref{KIC_data}. The procedure is as follows: We calculated luminosities and absolute brightnesses of the stars. Bolometric corrections were done in K-band using the closest grid point ($T_{eff}$,$\log g$) from published BC-coefficients of \cite{bessell}. For the flares we considered the spectral response of the Kepler photometer. 
Bolometric corrections were done after transforming $mag_{kep}$ into $V$ \citep{still}. Due to uncertainties of $R_{star}$ and $T_{eff}$, given in the KIC, the expected errors of these calculations are up to $85\:\%$. The values for bolometric flare luminosity and energy are presented in Tab. \ref{properties}. 
For the flares detected in the light curve of KIC10524994 we further investigated the correlation between the flare energy and the cadence of two neighbouring events. All the detected flares are superflares, according to their $1\sigma$ errors and the definition by \citet{schaefer}, which is at least 10 times the energy of the largest known solar event ($10^{32}\:$erg). For most of the superflares the waiting time is less than 100 days, but we could not find any dependence on the flares energies. 
%\begin{figure}
%\includegraphics[width=1\columnwidth]{flare_properties2.eps} 
 %\vspace{pt}
% \caption{The plot shows the waiting time over the flare energy. Due to the uncertainties of $T_{eff}$ and $R_{star}$, gathered by the KIC, the error of the luminosities can be up to $85\%$. Within $\pm 1 \sigma$ almost all the detected flares are superflares by definition in \citet{schaefer}.}
%  \label{flare_properties}
%\end{figure}
\subsection{Kepler Astrometry}
We used the TPF files of the targets to calculate their astrometric signal. For each time stamp the files provide a fully calibrated array of pixel fluxes, their $1\sigma$ uncertainties, information about cosmic ray incidences and a predicted motion of the targets calculated from the motion of a set of reference stars on the detector \citep{kinemuchi}. Monet et al. (2010) showed that the precision of a single measurement of a typical star can be as good as 0.001 pixel. {\bf The pixel scale of Kepler is about $4''/px$}. If one assumes the target stars are unresolved point sources, the signal of a flare would be well distributed over the optimal aperture of the target, hence the center of light would not be changed significantly by a flare. Note that it is not unusual for Kepler targets that a pixel mask is crowded by the flux of neighbouring stars. If flares are produced on such other sources, the center of light could change significantly towards the source of the brightening. Therefore we convolved the astrometric signal of the stars with the occurrence times of detected flares to test the hypothesis of a crowded field.

We used a simple center approximation \citep{howell} for the point-spread function (PSF) of the target to calculate the center of light and error bars for each time stamp for both detector axes using the calibrated pixel fluxes. After subtracting the predicted movement from the raw astrometric signal, the shift on the detector is mostly linearised. This movement might be a superposition of proper motion and trigonometric parallax, but can still contain a signal from binary interaction or differential velocity aberration \citep{monet}. We fitted a second order polynomial to the curve to eliminate this remaining trend.

\begin{figure}
\includegraphics[angle=270, width=\columnwidth]{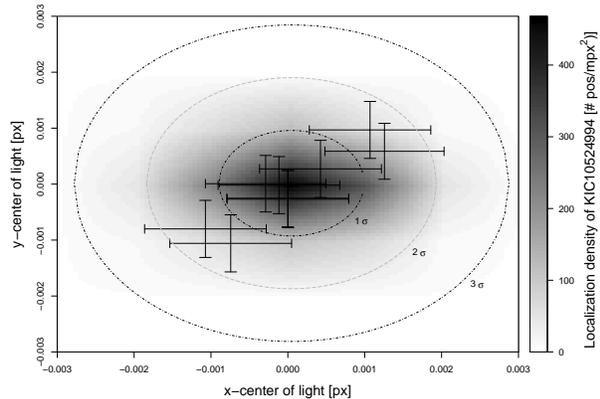} 
 %\vspace{pt}
 \caption{Localization density for KIC10524994 for quarter 0 after movement correction and trend subtraction. Black data points denote the center of light at the peak time $\pm 2\:$h for the largest flare at $96.06\:$d. {\bf The pixel scale is $\approx 4''/px$. The photometric barycenter has not changed significantly during the flare, which indicates the superflare to be originated on the target star.}}
  \label{center1}
\end{figure}

\begin{figure}
\includegraphics[angle=270, width=1\columnwidth]{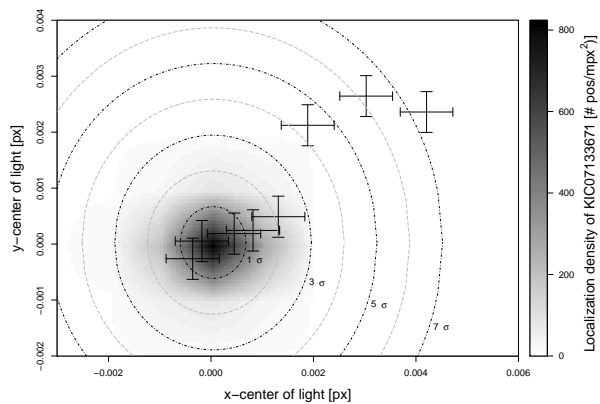} 
 %\vspace{pt}
 \caption{Localization density for KIC07133671 for quarter 0 after movement correction and trend subtraction. Black data points denote the center of light at the peak time $\pm2\:$h for the largest flare at $68.2\:$d. {\bf The pixel scale is $\approx 4''/px$.} The flare induces a shift of the photocenter of $\approx8\:\sigma$. The flare could have originated from a different star, located $>25\:$mas to the north east of the pointing target. It might be a background star or a binary companion.}
  \label{center2}
\end{figure}

{\bf Fig. \ref{center1} and Fig. \ref{center2} show a contour plot for the center of light for KIC10524994 and KIC07133671 on the detector. The grey levels represent the location density of the stars during quarter 0 in terms of the number of positions per square millipixel}. For each star we overplotted the astrometric signal at the peak time $\pm 2\:$h for the largest flare of each target (black data points) {\bf and the $\sigma$ levels for a fit of a two dimensional gaussian distribution of the stars localization (dotted ellipses)}. For KIC10524994 all the flare data are obviously within the bulk of data points around the quarterly averaged center of light, no more than $2\:\sigma$ away from it. For KIC07133671 3 data points are significantly outside the bulk of data points with up to $8\:\sigma$ distance from the averaged photometric barycenter. This is in good agreement with the photometric signal of that flare (Fig. \ref{flares}, lower right). From these results we conclude that the astrometric shift during the flare is a strong indication for an error source within the pixel mask of KIC07133671. This additional star causes a shift of the photocenter of $>25\:$mas to the north east of the pointing source, which is consistent with a projected separation of $> 22\:$AU.    

Note that we also checked the probability for a membership in any general galactic field for both targets from proper motion measurements \citep{roeser} of surrounding stars within $1 \:\text{deg}^2$. Given these proper motions, KIC10524994 ($\mu_{\alpha}=(-2.1\pm4.0)\:$mas/yr, $\mu_{\delta}=(-4.7\pm4.0)\:$mas/yr) can be considered as a general galactic field star. KIC07133671 ($\mu_{\alpha}=(-9.1\pm3.8)\:$mas/yr, $\mu_{\delta}=(-31.2\pm3.9)\:$mas/yr) significantly differs from the distribution of the surrounding stars. Since we have no indication for membership in any open clusters from the proper motion measurements, we assume that the targets are not younger than $1\:$Gyr, as this is the upper timescale over which a cluster is typically disrupted \citep{fuente}.  
 
\subsection{The frequency distribution of superflares for energies $> 10^{34}\:$erg}
\begin{figure}
\includegraphics[angle=270,width=1\columnwidth]{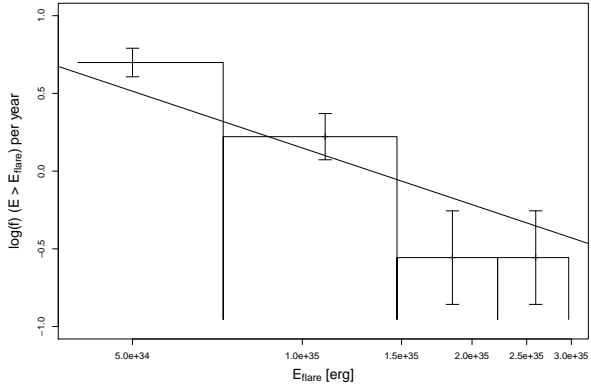} 
 %\vspace{pt}
 \caption{Cumulative frequency distribution for superflares of KIC10524994. The solid black line denotes the estimated power law approximation in this work ($\alpha \approx1.2$). It fits well to the frequency distribution of solar events \citep{schrijver}.}
  \label{energies}
\end{figure}

We used the energy distribution of the superflares of KIC10524994 to determine the cumulative frequency distribution $(f)$ for bolometric flare fluences above $10^{34}\:$erg. For this purpose we tried to fit with a power law $f \propto E^{-\alpha}$, as done by \citet{schrijver} for solar flares. Fig. \ref{energies} shows a histogram of the cumulative frequency distribution in units per time over the energy for superflares on the G-type star KIC10524994. The size of the energy bins is comparable to the averaged uncertainty of the superflare energies. With a Levenberg-Marquardt optimization routine we fitted a power-law with an index of $\alpha=1.2\pm0.2$ (see Fig. \ref{energies}, solid line). It is interesting to note, that this power-law index for KIC10524994 is consistent with the value estimated by \citet{schrijver} for solar flares of much lower energies ($10^{24}-10^{27}\:$erg).

\section{Discussion}
We could show that the frequency of superflares on KIC10524994 can be well described by a power-law index of $\alpha=1.2$. This is consistent with the power-law index of solar flares in a much lower energy range \citep{schrijver}. In order to improve the frequency statistics and to connect both energy ranges of the known solar flares and the superflares on sun-like stars by a certain power-law approximation, more sun-like stars need to be investigated with our detection and analysing method. Note that the superflare energies in this work are roughly 2-3 times larger than in \citet{maehara}, since \citet{maehara} did not include extinction in the energy calculations.
 
{\bf KIC10524994 and KIC07133671 were the only two presumable sun-like stars within an investigated sample of 14000 stars that show superflares with energies larger than $10^{35}\:$erg \citep{maehara}}. On the basis of only these two stars, \citet{maehara} estimated a frequency of one flare in $5000\:$yr for that energy range.

Since the photocenter of the emission is shifted only during the (only one observed) flare of the star KIC07133671, the flare did not originate on this solar-like star, but on a background or companion star.
Hence, KIC07133671 should not be included in the statistics of superflares of sun-like stars - reducing the rate of \citet{maehara} by a factor of two. Hence, the probability for the AD 774/5 event \citep{miyake,hambaryan} to be a solar superflare also is smaller than previously thought. The other presumable solar-like superflare star, KIC10524994, may be still a close binary with separation smaller than the Kepler PSF.

Kepler astrometry is a powerful tool to check whether a target pixel mask is crowded by error sources, e.g. background objects, that could be the origin for detected superflares. Though we only investigated the light curves of two stars, the rejection of KIC07133671 as a superflare star has a high impact on the frequency statistics of that energy range. This yields additional motivation for applying our method to the remaining superflare stars, probably to find more false positives, and hence, to further improve the frequency statistics. This is also necessary to find constraints for superflares on the sun. 

\section{Summary}

The light curves of two superflare stars, KIC07133671 and KIC10524994, were analysed to detect and characterize superflares. We further investigated the cyclic behaviour of the detected flares to check the hypothesis of possible companions indirectly. For KIC10524994 we could find a rotational period of $(11.87\pm 0.07)\:$d, which is compatible with \citet{maehara}. This period was compared to evolutionary models of the surface angular velocity. We could derive an age of $1.0^{+0.6}_{-0.4}\:$Gyr, which is only a rough estimate. With a Bayesian approach we could detect 19 flares, while 18 of them are new detections, compared to \citet{maehara}. For KIC07133671 we could find a period of $15.19\:$d, which is consistent with \citet{shiba}. We could find one superflare for KIC07133671, the one from \citet{maehara}. From the convolution of the astrometric signal of KIC07133671 with the occurrence time of its flare, there is high evidence that the flare is caused by a background star or a companion, indicating that the target should not be involved in the frequency statistics of superflares of solar twins. 

We determined the frequency of superflares in the energy range of $10^{34}-10^{35}\:$erg by a power law with an index of $\alpha=1.2\pm0.2$. A comparison to solar flares of much lower energies indicates that the distribution of solar flares and the distribution of superflares on KIC10524994 are assigned to the same power law index \citep{schrijver}. To improve the frequency statistics and to find constraints for superflares on the Sun, we plan to reanalyse all the sun-like stars within the population that \citet{maehara} investigated.

\section*{Acknowledgments}
We thank B. Schaefer and R. Arlt for fruitful suggestions and useful comments. MK and RN would like to thank DFG in SPP 1385 project NE515/34-1,34-2 for financial support. VVH would like to thank DFG in SFB TR7 for financial support. CG would like to thank DFG in project MU2695/13-1 for financial support. Finally, we would like to thank Donna Keeley for the language editing of our manuscript.

\appendix

\end{document}